\documentclass[lineno]{jfm}

\usepackage{graphicx}
\usepackage{newtxtext}
\usepackage{newtxmath}
\usepackage{natbib}
\usepackage{hyperref}
\usepackage{caption}
\hypersetup{
    colorlinks = true,
    urlcolor   = blue,
    citecolor  = black,
}

\newcommand{\RomanNumeralCaps}[1]

\newcommand{\la}{\left<}
\newcommand{\ra}{\right>}
\def\la{\left \langle}
\def\ra{\right \rangle}

\newcommand{\bU}{{\boldsymbol{{\cal U}}}}
\providecommand\bnabla{\boldsymbol{\nabla}}

\usepackage[normalem]{ulem}

\shorttitle{Reduced equation for gravity wave-mean flow interaction}
\shortauthor{B. Gallet, A. Tlili}

\title{Surface gravity wave–mean flow interaction with comparable spatial scales. Part I: reduced wave equations.}

\author{Basile Gallet  \corresp{\email{basile.gallet@cea.fr}} \& Alexandre Tlili
}

\affiliation{\aff{1}{Universit\'e Paris-Saclay, CNRS, CEA, Service de Physique de l'Etat Condens\'e, 91191 Gif-sur-Yvette, France.}}

\begin{document}
\maketitle


\begin{abstract}

We consider deep-water surface gravity waves propagating above a background flow whose spatial scale is comparable to the wavelength, focusing on the regime where the flow is slow compared to the wave velocity. 
We introduce an `equivalent solvability condition' method to construct reduced equations, demanding that, upon multiple-timescale expansion, the reduced equations share the same leading-order solution and first solvability condition as the original system. This approach turns the full 3D problem into a 2D reduced equation for the wave field. We derive such reduced equations for broad-band waves above a depth-invariant background flow, and for narrow-band waves above a fully 3D background flow. In the latter case the reduced equation takes the form of a Schr\"odinger equation involving the near-surface vorticity of the background flow only, with the impact of the near-surface horizontal flow divergence shown to be subdominant. Beyond the reduction in spatial dimensionality, the latter reduced equation describes the evolution of the wave field over the slow advective timescale of the background flow, thereby eliminating the computational burden of time-resolving the fast wave period. We illustrate the capabilities of the reduced equations through an analytical solution for the weak scattering of a wave packet by a patch of organized flow, followed by numerical solutions for stronger scattering of a wave packet by a patch of disorganized flow.

\end{abstract}


\section{Introduction}



Many fluid dynamics systems involve the interaction between a slowly evolving background flow and faster wave motion~\citep{buhler2014}. This is often because waves naturally arise in the system, be they inertial waves above a geostrophic flow in rotating turbulence, atmospheric Rossby waves above jet-like flows, internal gravity waves above vertically sheared flows, or Ocean swell and near-inertial waves (NIWs) propagating above mesoscale and submesoscale turbulence. Alternatively, in the laboratory the waves can be driven externally as a means to probe the background flow through sound-vorticity or surface-wave background-flow interaction~\citep{rayleigh1896theory,lund1989ultrasound,cerda1993interaction,berthet1995forward,labbe1998propagation,humbert2017surface,prabhudesai2022statistics}, or as a means to drive the flow, e.g. through acoustic streaming~\citep{nyborg1958acoustic,lighthill1978acoustic,vincent2024experimental,vincent2025phenomenology}.

In a similar fashion to geometric optics, a standard approach to characterize the effect of the background flow on the waves focuses on the regime of scale separation, where the wavelength is much shorter than the horizontal scale of the background flow. Then, through a WKB approach one obtains the ray-tracing equations, that is, a Hamiltonian set of equations for the trajectories of waves packets in the inhomogeneous background flow.  The properties of the waves field are then inferred from the computation of many rays~\citep{kunze1985near,gallet2014refraction,Boury2023}, or through the computation of direct solutions to the action density equation, using asympotic methods~\citep{boas2020directional,wang2023scattering,Wang2025} or prescriptions borrowed from equilibrium statistical mechanics~\citep{Tlili2025,tlili2026equilibrium}. Recently, a two-way coupled model between the surface waves and the background mean flow was derived by~\citet{vanneste2026consistent}, who combine the wave-action-density equation with the Craik-Leibovich equation, the latter encoding the feedback of the short waves onto the larger-scale background flow.

An obvious limitation of the ray-tracing approach is that systems of interest do not necessarily feature a separation between the wavelength and the scale of the background flow. For instance, various surface wave-mean flow  laboratory experiments operate in regimes where the wavelength is comparable to the horizontal scale of the background flow~\citep{teixeira2002distortion,vivanco2000surface,vivanco2004experimental,gutierrez2016surface,humbert2017wave}, and similarly for acoustic streaming systems~\citep{chini2014large,michel2019strong} and sound-vorticity interaction studies~\citep{berthet2003study}. In Earth's atmosphere, Rossby waves have a wavelength comparable to the planetary radius, leaving no room for scale separation with the mean flow. In the ocean, Langmuir cells are an important manifestation of surface-wave mean flow interactions, with a wavelength that is often comparable to the lateral extent of the cells. 

In the Physics literature, the regime of comparable scales is typically addressed in terms of resonant scattering, where incoming waves - particles, light, sound, etc -- get scattered at finite angle by a spatially localized scattering medium. A standard framework is the Born approximation, where the scattering effects are weak enough for the incoming wave to be considered unperturbed. The properties of the weak scattered waves are then computed perturbatively at large distance from the scattering medium (far-field regime). Obviously, such resonant-scattering problems are all the more easily addressed that the starting point is a simple wave equation, such as a Schr\"odinger equation or d'Alembert's equation. When it comes to deep-water surface waves, however, an immediate difficulty is that the governing equations  are notoriously more complicated than the above-mentioned simple equations. This perhaps explains why the comparable-scale regime has received far less attention than the scale-separation regime, with the notable exception of the studies by~\citet{phillips1959scattering} and~\citet{fabrikant1994influence}. 

The goal of the present study is to help close this knowledge gap by deriving simple equations to address the comparable-scale regime for surface waves above a background flow. Namely, we derive reduced equations governing the behavior of deep-water surface gravity waves above a weak background flow. The reduced equations are simple to interpret, typically taking  the familiar form of a Schr\"odinger equation. They are also much simpler to simulate numerically, as they turn the original fully 3D free-surface system into an equation for a 2D field evolving on a time scale much greater than the wave period. 

To some extent, the approach we propose is somewhat similar to -- and motivated by -- the celebrated study by~\citet{YBJ1997} (YBJ in the following), who derived a reduced equation governing the modulation of near-inertial waves (NIWs) by a background balanced flow. By offering a simplified framework, the YBJ equation has spawned many studies characterizing the interplay between NIWs and various background flows of interest. The YBJ approach has also been extended and applied to other wave systems~\citep{thomas2017new,wagner2017asymptotic}, leading to reduced equations elegantly obtained from the solvability conditions of weakly perturbed wave systems. In contrast with such previous studies, the present deep-water surface-wave problem includes one extra level of complexity: through the reduction approach, we aim at reducing the spatial dimension of the system from the original fully 3D system to a 2D reduced equation. For water waves above topography, \citet{thomas2018amplitude} address this issue by focussing on narrow-band waves around a central frequency and deriving a reduced equation using the asymptotic method of reconstitution~\citep{roberts1985introduction}. As we were finalizing the writing of the present article, we became aware of the recent preprint by~\citet{onuki2026reduced}, who also apply the method of reconstitution to narrow-band deep-water waves above background flows to derive a coupled model between the waves and the background flow. In contrast with such studies, in the present article we introduce an `equivalent solvability condition method', which allows us to address both narrow-band and broad-band wave fields. The method consists in constructing simpler, two-dimensional equations that share the same leading-order solution and the same solvability condition as the original fully 3D system. Another central aspect of the present work is our effort to cast the reduced equations in the simplest possible form, to ease physical interpretation and practical implementation. 


The structure of the article is the following. We first introduce the reduced equations, discussing their properties and conserved quantities. In section~\ref{sec:depthinvariant}, we introduce a reduced equation for broad-band deep-water surface gravity waves propagating above a slow, depth-invariant background flow. Then, in section~\ref{sec:NB3D}, we introduce a reduced equation governing narrow-band deep-water surface gravity waves propagating above a slow, fully 3D background flow. Sections~\ref{sec:full3D} through~\ref{sec:NB3Dderivation} are more technical and contain the derivation of the reduced equations using the equivalent solvability condition method. The reader uninterested in the mathematical details can skip directly to sections~\ref{sec:scatteringanalytical} and~\ref{sec:scatteringnumerical}, where we illustrate the possibilities offered by the reduced equations through analytical and numerical solutions.

\section{Reduced equation for broad-band waves above a depth-invariant background flow\label{sec:depthinvariant}}

\begin{figure}
    \centerline{\includegraphics[width=12 cm]{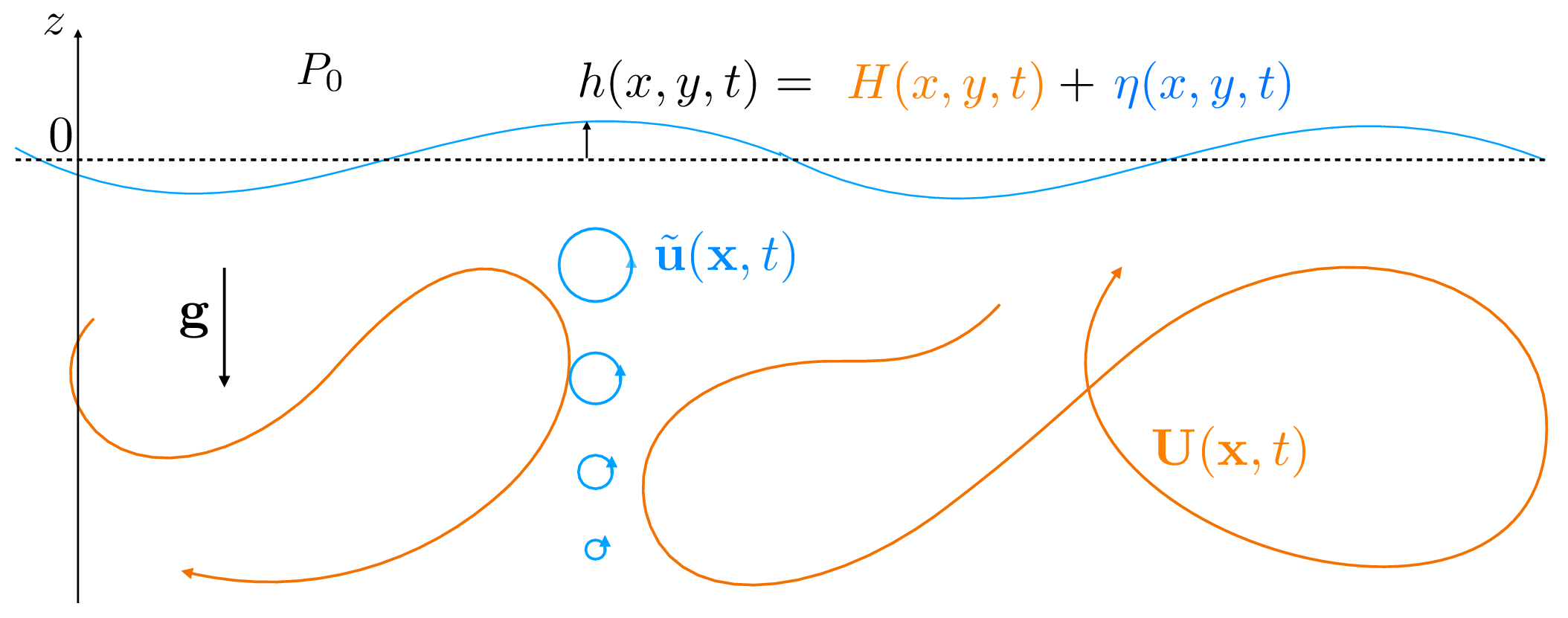} }
   \caption{\label{fig:schematic} Infinitesimal deep-water surface gravity waves above a background flow ${\bf U}({\bf x},t)$. The surface displacement $H(x,y,t)$ associated with the background flow has a negligible impact on the waves in the low-Froude number regime of interest here.}
\end{figure}

We consider the setup sketched in figure~\ref{fig:schematic}. An incompressible fluid of density $\rho$ occupies the half space $z\leq h(x,y,t)$, with periodic boundary conditions in the horizontal directions and gravity $g$ acting in the vertical direction. The fluid sits below an atmosphere of uniform pressure $P=P_0$ and negligible density. Neglecting viscous effects, the fluid motion is governed by the Euler equation for a divergence-free flow subject to gravity. The fluid hosts a 3D background flow ${\bf U}(x,y,z,t)$ associated with an interface deformation $H(x,y,t)$, together with infinitesimal deep-water surface gravity waves associated with a wavy velocity field $\tilde{\bf u}(x,y,z,t)=(\tilde{u},\tilde{v},\tilde{w})$ and a wavy surface displacement ${\eta}(x,y,t)$. Denoting as $L$ the characteristic horizontal scale of both the background flow of the wave field and as $U_0$ the characteristic velocity of the background flow, we focus on the low-Froude-number regime $U_0/\sqrt{gL}=\epsilon \ll 1$. That is, the background flow is much slower than the group velocity of the waves and its turnover time is much longer than the wave period. We derive reduced equations governing the evolution of the wave field in two regimes of interest: broad-band waves above a depth-invariant background flow, and narrow-band waves above a depth-dependent background flow. 

To characterize the wave field, the quantities of interest are typically the wavy surface displacement $\eta(x,y,t)$, and/or the wavy vertical velocity evaluated at the surface, $w_s(x,y,t)$. The reduced equations are conveniently cast in terms of the complex variable:
\begin{align}
\chi(x,y,t)=\frac{1}{\sqrt{2}}D^{-\frac{1}{4}} \eta + \frac{i}{\sqrt{2}} D^{-\frac{3}{4}} w_s,
\label{eq:chi_definition}
\end{align}
where the operator $D$ corresponds to multiplication by the horizontal wavenumber $k$ in spectral space. In equation~(\ref{eq:chi_definition}) and in the following, we have non-dimensionalized all variables using the spatial scale $L$ of the background flow (comparable to the wavelength) and the time scale $\sqrt{L/g}$ (see equation~(\ref{eq:adim})). 
The variable $\chi$ naturally arises in the weak-turbulence treatment of surface gravity waves~\citep{zakharov_1982,zlf_book,Nazarenko2011}.


We first consider the case of a depth-invariant background flow, ${\bf U}={\bf U}_\perp=[U(x,y,t),V(x,y,t),0]$. Such an incompressible flow stems from a streamfunction, ${\bf U}=-\bnabla \times [\Psi(x,y,t) {\bf e}_z]$. The reduced equation reads:
 \begin{align}
 \partial_t \chi + i D^{{\frac{1}{2}}} \chi + J(\Psi,\chi) +\frac{1}{4}  D^{-1} \{ J(\Delta \Psi,  D^{-1} \chi)   \}  & =  0 \, .  \label{eq:sec2reducedBB}
 \end{align}
The second term of this equation encodes the dispersion relation of deep-water surface gravity waves, and the third and fourth terms encode the effect of the background flow. While the third term is a standard advective term, the fourth term involves the nonlocal operator $D$.  As compared to the original 3D system, the reduction in complexity achieved by equation~(\ref{eq:sec2reducedBB}) is two-fold. Firstly, equation~(\ref{eq:sec2reducedBB}) involves only two spatial dimensions. Secondly, while equation~(\ref{eq:sec2reducedBB}) still involves fast-wave dynamics through the dispersion term $ i D^{{\frac{1}{2}}} \chi$, at the computational level this term can be integrated exactly~\citep{lawson1967generalized}, allowing one to use large time steps comparable to the turnover time of the background flow.
From the solution to equation~(\ref{eq:sec2reducedBB}), the wavy interfacial displacement and surface velocity are retrieved as:
 \begin{align}
\eta & =\frac{1}{\sqrt{2}} (D^{{\frac{1}{4}}} \chi + D^{{\frac{1}{4}}} \chi^*) \, , \label{eq:sec2etavschi}\\
w_s & =- \frac{i}{\sqrt{2} } (D^{{\frac{3}{4}}} \chi - D^{{\frac{3}{4}}} \chi^*) \, . \label{eq:sec2wvschi}
 \end{align}
As shown in the following, an appealing aspect of equation (\ref{eq:sec2reducedBB}) is that it conserves wave action exactly. When the background flow is steady, equation (\ref{eq:sec2reducedBB}) also conserves energy exactly.

\subsection{Action conservation\label{sec:actioncons}}

Multiply equation~(\ref{eq:sec2reducedBB}) by $\chi^*$ before adding the complex conjugate and averaging over space. Denoting horizontal area average as $\la \cdot \ra_{\bf x}$ and making extensive use of the formula $\la f^* D^{\alpha} \{g \} \ra_{\bf x} = \la D^{\alpha} \{ f^* \} g \ra_{\bf x}$ (easily obtained by expanding $f$ and $g$ in Fourier series), one obtains:
 \begin{align}
\frac{\mathrm{d}}{\mathrm{d}t} \la |\chi|^2 \ra_{\bf x} & = 0 \, .
 \end{align}
We conclude that the wave action ${\cal A}=\la |\chi|^2 \ra_{\bf x}$ is conserved. 
This conservation law results from the separation of timescales between the fast wave period and the slow turnover time of the background flow.

\subsection{Energy conservation}

If the background flow is steady, the system is invariant to translations in time and therefore it conserves energy. The energy functional is easily identified after multiplying equation~(\ref{eq:sec2reducedBB}) with $i$ to  recast it under the form of a Schr\"odinger equation, $i \partial_t \chi= {\cal H}\{\chi \}$, where the (hermitian) Hamiltonian operator is:
 \begin{align}
\nonumber  {\cal H}\{\chi \} & = D^{{\frac{1}{2}}} \chi -i J(\Psi,\chi)-\frac{i}{4}D^{-1} \{ J(\Delta \Psi, D^{-1} \chi) \} \, .
 \end{align}
 The conserved energy is then simply:
  \begin{align}
E & = \la \chi^* {\cal H}\{\chi \} \ra_{\bf x} \\
& = \la |D^{{\frac{1}{4}}} \chi|^2  - i \Psi J(\chi,\chi^*) - i \, \frac{\Delta \Psi}{4} J \left( D^{-1}\chi, D^{-1}\chi^* \right) \ra_{\bf x} \, ,  \label{eq:sec2energyinvariant}
 \end{align}
 where we have performed a few integrations by parts to highlight the symmetric roles of $\chi$ and $\chi^*$.

\section{Reduced equation for narrow-band waves above a depth-dependent background flow\label{sec:NB3D}}

We can also derive a reduced equation for the more general case of a depth-dependent background flow ${\bf U}(x,y,z,t)$. However, in this case we restrict attention to narrow-band wave fields: the frequency content of the wave field is peaked around a dominant dimensional angular frequency $\omega_0$ with a narrow, ${\cal O}(\epsilon)$ relative bandwidth around this central frequency. It then proves convenient to non-dimensionalize the equations using the timescale $1/\omega_0$ and the spatial scale $g/\omega_0^2$. The dimensionless central frequency is equal to one, corresponding to a unit dimensionless wavenumber. In a somewhat similar fashion to the YBJ approach, the reduced equation is cast in terms of a demodulated complex wave amplitude ${M}(x,y,t)$ that absorbs the fast oscillation at the central frequency:
\begin{align}
{M}(x,y,t)=\chi(x,y,t) e^{it} (1+{\cal O}(\epsilon)) \, , \label{eq:vaguedefcalM}
\end{align}
where the small, ${\cal O}(\epsilon)$ term in the parentheses is carefully chosen to obtain a reduced equation that only features a horizontally non-divergent effective flow (see section~\ref{sec:eliminating} and equation~(\ref{eq:defcalM}) for a precise definition of ${M}$). The reduced equation reads:
\begin{align}
\partial_t {M} + J(\psi,{M}) - \frac{i}{4} \left( \Delta {M} + {M} \right) & = 0 \, , \label{eq:sec2calMstandardadim}
\end{align}
where the second term describes advection by an effective, solenoidal 2D flow, and the third term stems from an expansion of the intrinsic dispersion relation of deep-water surface gravity waves around the central frequency. 
We stress the fact that equation~(\ref{eq:sec2calMstandardadim}) must be integrated from an initial condition for ${M}$ that corresponds to narrow-band waves. The initial condition should have significant Fourier amplitudes for dimensionless wavectors of magnitude $k=1+{\cal O}(\epsilon)$ only, which indeed corresponds to narrow-band waves with dimensionless frequencies $1+{\cal O}(\epsilon)$.

Beyond the reduction in spatial dimension, an appealing aspect of equation~(\ref{eq:sec2calMstandardadim}) is that ${M}$ evolves at the slow, ${\cal O}(\epsilon)$ turnover frequency of the background flow and at the small ${\cal O}(\epsilon)$ detuning of the waves around the central frequency, while ${M}$ does not evolve at the fast, ${\cal O}(1)$ wave frequency. At the practical level, this means that one can numerically integrate equation~(\ref{eq:sec2calMstandardadim}) using timesteps much greater than the wave period. 

The effective, two-dimensional streamfunction $\psi$ entering equation~(\ref{eq:sec2calMstandardadim}) is given by:
\begin{align}
\psi(x,y,t) & = \left(\Delta^{-1}+\frac{1}{4} \right) \left\{  \int_{-\infty}^0 2 e^{2z} \zeta(x,y,z,t) \mathrm{d}z  \right\} \, , \label{eq:defpsi3D}
\end{align}
where $\zeta=(\bnabla \times {\bf U})\cdot {\bf e}_z$ denotes the vertical vorticity of the background flow.
To obtain equation~(\ref{eq:defpsi3D}) we have assumed that the effective flow has a vanishing horizontal area average over the domain, deferring the inclusion of horizontally uniform vertically sheared background flows to appendix \ref{app:uniform}. We recover the result of~\citet{fabrikant1994influence} that only the vertical vorticity of the background flow acts to scatter surface gravity waves. 
By contrast, the horizontal divergence of the near-surface flow has a subdominant impact on the waves compared to the near-surface vertical vorticity, as established in section~\ref{sec:eliminating}.

From the solution to equation~(\ref{eq:sec2calMstandardadim}) and the definition~(\ref{eq:vaguedefcalM}), the leading-order wavy interfacial displacement and vertical velocity at the surface are obtained as:
 \begin{align}
\eta(x,y,t) & = \frac{1}{\sqrt{2}} {M}(x,y,t)e^{-it} + \text{c.c.} \, , \label{eq:sec2etavscalM}\\
w_s(x,y,t) & = -\frac{i}{\sqrt{2}} {M}(x,y,t) e^{-it} + \text{c.c.}   \, ,  \label{eq:sec2wvscalM}
 \end{align}
up to corrections of higher order in $\epsilon$. To obtain these expressions, we have used the fact that the operator $D=1+{\cal O}(\epsilon)$ under the narrow-band approximation.

Equations~(\ref{eq:vaguedefcalM}), (\ref{eq:defpsi3D}), (\ref{eq:sec2etavscalM}) and~(\ref{eq:sec2wvscalM}) hold provided the background flow has a scale comparable to the wavelength. The situation where the background flow also includes much larger structures is deferred to the appendix~\ref{app:scaleseparation}, focused on the limit of scale separation. In the appendix, we first consider the situation where the background flow includes a horizontally uniform vertically sheared component. This situation is easily dealt with by 
augmenting equation~(\ref{eq:sec2calMstandardadim}) with a term encoding advection by the horizontally averaged effective background flow. 
Additionally, in the same appendix we consider the situation where the effective background flow includes extended horizontally divergent structures over a large, ${\cal O}(1/\epsilon)$ spatial scale. We obtain that equations~(\ref{eq:vaguedefcalM}), (\ref{eq:sec2etavscalM}) and (\ref{eq:sec2wvscalM}) are supplemented by a phase factor that slowly varies with space and time, see equations (\ref{eq:calMSS}-\ref{eq:wSS}). Such a phase factor does not affect the typical metrics used to quantify the wave field, such as the local action density $|{M}|^2$, or the local variance of surface elevation averaged over one wave period and the local variance of interfacial slope averaged over one wave period, which are respectively given by $|{M}|^2$ and $|\bnabla {M}|^2$ to leading order, see equations~(\ref{eq:dispvariance}) and (\ref{eq:slopevariance}). These metrics are readily obtained from a solution of equation~(\ref{eq:sec2calMstandardadim}).


As in section~\ref{sec:depthinvariant}, an appealing aspect of the reduced equation (\ref{eq:sec2calMstandardadim}) is that it exactly conserves wave action and, when the background flow is steady, it exactly conserves energy.

\subsection{Action conservation\label{sec:actioncons3D}}

Equation~(\ref{eq:sec2calMstandardadim}) being a Schr\"odinger equation, it exactly conserves wave action, the action density $|{M}|^2$ being analogous to the probability of presence of a quantum particle. A local version of this conservation law is readily obtained by multiplying equation~(\ref{eq:sec2calMstandardadim}) with ${M}^*$ and adding the complex conjugate, which leads to:
\begin{align}
\partial_t \left(|{M}|^2 \right) + \bnabla \cdot \left[ - |{M}|^2 \bnabla \times (\psi {\bf e}_z) + \frac{i}{4} \left( {M} \bnabla {M}^* - {M}^* \bnabla {M} \right) \right] & = 0 \, .
\end{align}
We recognize a conservation equation, the time evolution of the action density evolving as a result of the divergence of a flux of wave action. The divergence vanishes upon space average, which leads to the conservation of the overall wave action ${\cal A}=\la |{M}|^2 \ra_{\bf x}$.

\subsection{Energy conservation\label{sec:energycons3D}}

If the background flow is steady, equation~(\ref{eq:sec2calMstandardadim}) also conserves an energy invariant. We identify the latter by mutliplying the equation by $i$ to highlight the Schr\"odinger structure of the equation, $i \partial_t {M}={\cal H}\{{M} \}$ with ${\cal H}$ a Hamiltonian operator, before computing the energy invariant as $E= \la {M}^* {\cal H}\{{M} \}\ra_{\bf x}$. After a few integrations by parts this leads to the invariant:
\begin{align}
E & = \la -i \psi J({M},{M}^*) + \frac{1}{4} |\bnabla {M}|^2 - \frac{1}{4} | {M}|^2 \ra_{\bf x}  \\
& = \la -i \psi J({M},{M}^*) + \frac{1}{4} |\bnabla {M}|^2\ra_{\bf x}  - \frac{{\cal A}}{4} \, .
\end{align}
Because ${\cal A}$ is conserved on its own, the simpler invariant $\la -i \psi J({M},{M}^*) + \frac{1}{4} |\bnabla {M}|^2 \ra_{\bf x}$ is exactly conserved.

\subsection{Sanity check: narrow-band waves above a depth-invariant background flow}

For a $z$-invariant background flow stemming from a streamfunction $\Psi(x,y,t)$, we have $\int_{-\infty}^0 2 e^{2z} \zeta(x,y,t) \mathrm{d}z =\zeta(x,y,t) =\Delta \Psi$. Substituting into~(\ref{eq:defpsi3D}) yields:
\begin{align}
\psi & = \Psi + \frac{\Delta \Psi}{4} \, ,
\label{eq:psieff_barotropic}
\end{align}
so that the reduced equation~(\ref{eq:sec2calMstandardadim}) becomes:
 \begin{align}
\partial_t {M} + J\left(\Psi+\frac{\Delta \Psi}{4}, {M} \right)  - \frac{i}{4} \left( \Delta {M} + {M} \right) & = 0 \, . \label{eq:singlefreq}
\end{align}
This equation governs the evolution of narrow-band waves over a depth-invariant background flow, and as such it can also be obtained as the narrow-band limit of equation~(\ref{eq:sec2reducedBB}).

\section{The full 3D problem: deep-water surface gravity waves over a slow background flow\label{sec:full3D}}


We now derive the reduced equations from the full 3D system sketched in figure~\ref{fig:schematic}. The Euler equation for the full 3D divergence-free velocity field ${\bf u}(x,y,z,t)$ reads:
\begin{align}
\partial_t {\bf u} + ({\bf u} \cdot \bnabla) {\bf u} =- \bnabla p \, , \qquad \bnabla \cdot {\bf u} = 0 \, ,
\end{align}
where we have introduced the reduced pressure $p=(P-P_0)/\rho + gz$.
Denoting with a subscript $\cdot |_h$ quantities evaluated at $z=h(x,y,t)$ and with a subscript $\perp$ the horizontal components, the kinematic boundary condition reads:
\begin{align}
\partial_t \eta + {\bf u}_{\perp}|_h \cdot \bnabla_\perp h = w|_h \, ,
\end{align}
while the dynamic boundary condition $P|_h=P_0$ reduces to:
\begin{align}
p|_h =g h \, .
\end{align}


\subsection{Non-dimensionalization}

We denote as $L$ the typical wavelength of the wave field, which is also the characteristic horizontal scale of the background flow, and we non-dimensionalize the equations using $L$ and $g$:
\begin{align}
{\bf u}= \sqrt{g L} {\bf u}^\# \, , \qquad h=L {h}^\# \, , \qquad t=\sqrt{\frac{L}{g}} {t}^\# \, , \qquad p=g L {p}^\# \, , \qquad {\bf x}= L {{\bf x}}^\# \, , \qquad \Delta= {\Delta}^\#/L^2 \, . \label{eq:adim}
\end{align}
Dropping the $\#$ for brevity, the dimensionless equations read:
\begin{align}
\partial_t {\bf u} + ({\bf u} \cdot \bnabla) {\bf u} & =- \bnabla p \, , \label{eq:Euleradim}\\
\partial_t h + {\bf u}_{\perp}|_h \cdot \bnabla_\perp h & = w|_h \, \label{eq:kinematicadim}\\
p|_h & = h \, . \label{eq:dynamicadim}
\end{align}
To focus on the wave dynamics, we work mainly with the divergence of the Euler equation, together with the vertical component of the Euler equation evaluated at the surface:
\begin{align}
\partial_t h + {\bf u}_{\perp}|_h \cdot \bnabla_\perp h & = w|_h \, , \\
p|_h &  = h \, , \\
\Delta p & = - \bnabla \cdot [({\bf u} \cdot \bnabla) {\bf u}] \, , \\
\partial_t w|_h + ({\bf u} \cdot \bnabla w)|_h & = -(\partial_z p)|_h \, . \label{eq:wadimnolin}
\end{align}
%

\subsection{Linearization around a slow background flow}

We consider a background flow at the scale $L$ of the wavelength, with a characteristic speed $U_0$ much slower than the group velocity of the waves. That is, the Froude number of the background flow is small, $\epsilon= U_0/\sqrt{gL} \ll 1$, providing the small parameter of the asymptotic expansion. The turnover time of such a slow background flow is $L/U_0 \gg \sqrt{L/g}$, and therefore the background flow varies with the slow time variable $T=\epsilon t$ only.
We thus denote the dimensionless velocity, surface elevation and pressure of the background flow as ${\bf U}({\bf x},T)$, $H(x,y,T)$ and ${\cal P}({\bf x},T)$, respectively. 
We estimate the magnitudes of $H(x,y,T)$ and ${\cal P}({\bf x},T)$ using equations~(\ref{eq:Euleradim}-\ref{eq:dynamicadim}) in the absence of surface waves (see~\citet{phillips1959scattering} and~\citet{fabrikant1994influence} for similar estimates). Inserting  ${\bf U}({\bf x},T)={\cal O}(\epsilon)$ into (\ref{eq:Euleradim}) yields ${\cal P}={\cal O}(\epsilon^2)$, which after substitution into~(\ref{eq:dynamicadim}) yields $H={\cal O}(\epsilon^2)$. Finally, equation~(\ref{eq:kinematicadim}) provides an expression for the vertical velocity at the free surface:
\begin{align}
W|_H = \epsilon \partial_T H + {\bf U}_\perp|_H \cdot \bnabla H = {\cal O}(\epsilon^3)  \label{eq:BCWfull}
\end{align}
At the order of accuracy of the following expansion it is sufficient to expand this boundary condition into an effective no-penetration boundary condition, with the vertical flow satisfying:
\begin{align}
W|_0 &= 0  \, , \,\label{eq:BCWexpanded} \\
(\partial_z W)|_0 &= - \bnabla_\perp \cdot {\bf U}_\perp|_0  \, ,
\end{align}
where~(\ref{eq:BCWexpanded}) holds up to negligible ${\cal O}(\epsilon^2)$ terms. 

We now consider weak waves of typical (dimensional) wavelength $L$ and slope $\delta \ll \epsilon \ll 1$ propagating on the base flow. We focus on linear wave dynamics above the background flow and we linearize the equations accordingly. In doing so, we keep terms of order $\epsilon \delta$, while neglecting terms of order both $\delta^2$ and $\epsilon^2 \delta$. This ordering allows us to neglect the deformation $H(x,y,T)$ of the interface associated with the base state (see \citet{zuccoli2025deep} for a situation where the background surface deformation is significant). We thus replace quantities evaluated at $z=H(x,y,T)$ by the same quantities evaluated at $z=0$, up to negligible ${\cal O}(\epsilon^2)$ corrections. This procedure leads to:
\begin{align}
\partial_t \eta -w_s & =- {\bf U}_{\perp}|_0 \cdot \bnabla_\perp \eta - \eta (\bnabla_\perp \cdot {\bf U}_\perp|_0) &+{\cal O}(\epsilon^2 \delta, \delta^2) \, , \label{eq:adimscaled13D}\\
\eta-\tilde{p}|_0 & = 0 &+{\cal O}(\epsilon^2 \delta, \delta^2) \, ,  \label{eq:adimscaled23D}\\
\Delta \tilde{p} & = - \bnabla \cdot [(\tilde{{\bf u}} \cdot \bnabla) {\bf U} + ({\bf U} \cdot \bnabla) {\tilde{\bf u}}] & + {\cal O}(\delta^2) \, , \label{eq:adimscaled33D}\\
\partial_t w_s + (\partial_z \tilde{p})|_0 & = - {\bf U}_\perp|_0 \cdot \bnabla_\perp w_s  +  (\bnabla_\perp \cdot {\bf U}_\perp|_0) w_s &+{\cal O}(\epsilon^2 \delta, \delta^2) \, , \label{eq:adimscaled43D}
\end{align}
where we  denote the wavy velocity departure from the base state as $\tilde{\bf u}=(\tilde{u},\tilde{v},\tilde{w})$, the wavy reduced pressure as $\tilde{p}$, the wavy surface displacement as $\eta$, and the wavy vertical velocity at the surface as $w_s(x,y,t)=\tilde{w}|_0$.



\section{Derivation of the reduced equation for a depth-invariant background flow}

To illustrate the method on a slightly simpler situation, we first consider a depth-invariant background flow ${\bf U}={\bf U}_\perp=[U(x,y,T),V(x,y,T),0]=-\bnabla \times [\Psi(x,y,T) {\bf e}_z]$, where $\Psi$ denotes the streamfunction. Using the incompressibility constraint $\bnabla_\perp \cdot {\bf U}_\perp=0$, equations~(\ref{eq:adimscaled13D}-\ref{eq:adimscaled43D}) reduce to:
\begin{align}
\partial_t \eta -w_{s} & =- {\bf U}_{\perp} \cdot \bnabla_\perp \eta  &+{\cal O}(\epsilon^2 \delta, \delta^2) \, , \label{eq:adimscaled1}\\
\eta-\tilde{p}|_0 & = 0 &+{\cal O}(\epsilon^2 \delta, \delta^2) \, ,  \label{eq:adimscaled2}\\
\Delta \tilde{p} & = - \bnabla \cdot [(\tilde{{\bf u}} \cdot \bnabla) {\bf U} + ({\bf U} \cdot \bnabla) {\tilde{\bf u}}] & + {\cal O}(\delta^2) \, , \label{eq:adimscaled3}\\
\partial_t w_s + (\partial_z \tilde{p})|_0 & = - {\bf U}_\perp \cdot \bnabla_\perp w_s  &+{\cal O}(\epsilon^2 \delta, \delta^2) \, . \label{eq:adimscaled4}
\end{align}

We consider the equations above in a domain $[0,{\cal L}]\times[0,{\cal L}] \times (-\infty,0]$ with periodic boundary conditions in the horizontal. We seek a solution under the following multiple-timescale form:
\begin{align}
\eta & =\delta [\eta_0(x,y,t,T) + \epsilon  \eta_1(x,y,t,T) + \dots] \, , \label{eq:expansion}\\
\tilde{p} & =\delta [p_0({\bf x},t,T) + \epsilon  p_1({\bf x},t,T) + \dots] \, , \\
\tilde{{\bf u}} & =\delta [{\bf u}_0({\bf x},t,T) + \epsilon  {\bf u}_1({\bf x},t,T) + \dots] \, , \\
w_s& =\delta [w_{s0}(x,y,t,T) + \epsilon w_{s1}(x,y,t,T)  + \dots] \, . \\
\end{align}
where the slow time variable is $T=\epsilon t$.
Finally, in line with the low-Froude-number assumption we scale the background flow as:
\begin{align}
{\bf U}=\epsilon {\bf U}_1(x,y,T) \, , \qquad  \Psi=\epsilon \Psi_1(x,y,T) \, . \label{eq:scalingBGflow}
\end{align}

\subsection{Lowest order, ${\cal O}(\delta)$ \label{sec:lowestorder}}

To order $\delta$, equations~(\ref{eq:adimscaled1}-\ref{eq:adimscaled4}) yield:
 \begin{align}
\partial_t \eta_0 - w_{s0} & = 0 \, , \label{eq:kinOzero} \\
\eta_0-{p_0}|_0 & = 0  \, , \label{eq:dynOzero}\\
\Delta p_0 & = 0 \, \label{eq:LappOzero}\\
 \partial_t w_{s0} + (\partial_z p_0)|_0 & = 0 \label{eq:EulerwOzero}\, . 
 \end{align}
 We recognize the equations governing standard linear potential surface waves. In spectral space, the solution to equation~(\ref{eq:LappOzero}) that vanishes for $z \to -\infty$ is:
 \begin{align}
p_0({\bf x},t,T) & = \Sigma_{{\bf k}\in {\cal D}} {s}_{\bf k}(t,T) e^{i {\bf k}\cdot {\bf x}} e^{kz} \, , \label{eq:solp}
 \end{align}
 where $k=|{\bf k}|$ and the sum is over the ensemble ${\cal D}$ of horizontal wavevectors compatible with the periodic boundary conditions: ${\bf k}=(n_x,n_y,0) \times (2 \pi)/{\cal L}$ with $(n_x,n_y) \in \mathbb{Z}^2$. 
From equation~(\ref{eq:solp}) we obtain $(\partial_z p_0)|_0=D\{p_0|_0 \} = D\{\eta_0\}$, where the operator $D$ corresponds to multiplication by wavenumber $k$ in spectral space, and we have used equation~(\ref{eq:dynOzero}) to obtain the last equality.
We can thus recast equations (\ref{eq:kinOzero}) and (\ref{eq:EulerwOzero}) as:
 \begin{align}
\partial_t \eta_0 - w_{s0} & = 0 \, , \label{eq:eta0full} \\
 \partial_t w_{s0} + D\{ \eta_0 \} & = 0 \, . \label{eq:w0full} 
 \end{align}
These are the standard equations governing 2D linear potential waves (for instance, they correspond to the linearized version of the higher-order spectral (HOS) equations, see~\citet{dommermuth1987high,west1987new,onorato2002freely,dyachenko2004weak,zhang2022numerical,higgins2024numerical}). These equations lead to the standard dispersion relation of surface gravity waves, under the dimensionless form $\omega_k=\sqrt{k}$ with $\omega_k$ the dimensionless angular frequency (we describe the entire wavefield considering positive frequencies only and all the wavevectors ${\bf k}\in {\cal D}$). The solution to equations~(\ref{eq:eta0full}-\ref{eq:w0full}) is:
\begin{align}
\eta_0 & = \Sigma_{{\bf k}\in {\cal D}} A_{\bf k}(T) e^{i ({\bf k}\cdot {\bf x}-\omega_k t)} +\text{c.c.}\, , \label{eq:soleta0}\\
w_{s0} & = -i \Sigma_{{\bf k}\in {\cal D}} \omega_k A_{\bf k}(T) e^{i ({\bf k}\cdot {\bf x}-\omega_k t)} +\text{c.c.} \, , \label{eq:solw0}
 \end{align}
where the complex coefficients $A_{\bf k}$ depend on the slow time variable $T$. 

The lowest-order wavy velocity field ${\bf u}_0$ in the interior of the fluid domain is easily deduced from the ${\cal O}(\delta)$ vorticity equation, $\partial_t (\bnabla \times {\bf u}_0) = {\bf 0}$. We satisfy the latter equation by considering a potential flow solution, ${\bf u}_0 = \bnabla \phi$. The incompressibility constraint then yields $\Delta \phi = 0$, and using the boundary condition~(\ref{eq:solw0}) on $(\partial_z \phi) |_0 = w_{s0}$ we obtain the expression of the velocity potential $\phi({\bf x},t,T)$:
 \begin{align}
\phi & = -i \Sigma_{{\bf k}\in {\cal D}} \frac{A_{\bf k}(T)}{\omega_k} e^{i ({\bf k}\cdot {\bf x}-\omega_k t)} e^{kz} +\text{c.c.} \, . \label{eq:solphi0}
 \end{align}

\subsection{Next order, ${\cal O}(\epsilon \delta)$, and solvability condition}

To order $\epsilon \delta$, equations~(\ref{eq:adimscaled1}-\ref{eq:adimscaled4}) yield:
 \begin{align}
\partial_t \eta_1 - w_{s1} & = R_1 \, , \label{eq:kinOone} \\
\eta_1-{p_1}|_0 & = R_2  \, , \label{eq:dynOone}\\
\Delta p_1 & = R_3 \, \label{eq:LappOone}\\
 \partial_t w_{s1} + (\partial_z p_1)|_0 & = R_4 \label{eq:EulerwOone}\, ,
 \end{align}
where the rhs terms read:
 \begin{align}
R_1 & = - \partial_T \eta_0 - {\bf U}_1 \cdot \bnabla \eta_0 \, , \\
R_2 & = 0 \, , \\
R_3 & = - \bnabla \cdot [({{\bf u}_0} \cdot \bnabla) {\bf U}_1 + ({\bf U}_1 \cdot \bnabla) {{\bf u}_0}] \, , \\
R_4 & =  - \partial_T w_{s0} - {\bf U}_1 \cdot \bnabla w_{s0} \, .
 \end{align}

Denoting by $\la \cdot \ra$ an average over $x$, $y$ and the fast time variable $t$, we consider the equation resulting from  the linear combination $\la e^{-i({\bf k}\cdot {\bf x}-\omega_k t)} [ \right.$(\ref{eq:EulerwOone}) $+\partial_t$ (\ref{eq:kinOone}) $+k$ (\ref{eq:dynOone}) $-\int_{-\infty}^0 e^{k {z}}$ (\ref{eq:LappOone}) $\mathrm{d} {z} \left.]\ra$. 
After multiple integrations by parts, one obtains that the lhs of the resulting equation vanishes. We conclude that the rhs of the equation must vanish as well, which provides the solvability condition (see e.g. \citet{thomas2018amplitude} for a similar solvability condition). Denoting vertical integration over $z\in (-\infty,0]$ with an overbar, that is $\overline{\cdot}=\int_{-\infty}^0 \cdot \, \mathrm{d}z$, this solvability condition reads:
\begin{align}
\la e^{-i({\bf k}\cdot {\bf x}-\omega_k t)} \left(R_4+\partial_t R_1 + k R_2 -  \overline{e^{k {z}} R_3} \right) \ra & = 0 \, . \label{eq:rawSC}
\end{align}
In appendix~\ref{app:SCzinv}, we express the contribution from $R_3$ in terms of the wavy vertical velocity evaluated at the surface, the end result being:
\begin{align}
\la e^{-i({\bf k}\cdot {\bf x}-\omega_k t)}  \overline{ e^{k {z}} R_3 } \ra & =  \frac{1}{2k^2} \la e^{-i ({\bf k}\cdot {\bf x}-\omega_k t)}   J(\Delta \Psi_1,   w_{s0} )  \ra \, .
\end{align}

Substitution into~(\ref{eq:rawSC}) yields:
\begin{align}
\la e^{-i({\bf k}\cdot {\bf x}-\omega_k t)} \left[ - \partial_T w_{s0} - {\bf U}_1 \cdot \bnabla w_{s0}+i \omega_k ( \partial_T \eta_0 + {\bf U}_1 \cdot \bnabla \eta_0) - \frac{1}{2k^2}   J(\Delta \Psi_1, w_{s0}) \right] \ra & = 0 \, . \label{eq:cleanerSC}
\end{align}
Upon inserting the expressions~(\ref{eq:soleta0}-\ref{eq:solw0}), the solvability condition~(\ref{eq:cleanerSC}) provides a set of coupled nonlinear first-order ODEs governing the slow-time dependence of the coefficients $\{A_{\bf k}(T)\}_{\bf k}$. This shows that the solvability condition fully determines the slow evolution of the leading-order wavefield. However, the resulting infinite set of ODEs appears cumbersome to deal with. We thus follow a different route in the following: we seek a set of simpler, 2D equations that leads to the same solution at ${\cal O}(\delta)$ and the same solvability condition at ${\cal O}( \epsilon \delta)$ as the original 3D system.

\section{Reduced, two-dimensional equations}

Our starting point is the following ansatz for the structure of the reduced, 2D equations:
 \begin{align}
\partial_t \eta  - w_s & = R_5 \, , \label{eq:2Deqeta}\\
 \partial_t w_s  + D\{ \eta \} & = R_6 \, , \label{eq:2Deqw}
 \end{align}
where $R_5$ and $R_6$ are ${\cal O}(\epsilon \delta)$ terms that will be specified later. Following the same multiple-timescale expansion as~(\ref{eq:expansion}), at ${\cal O}(\delta)$ we obtain the same equations~(\ref{eq:eta0full}-\ref{eq:w0full}), with the same solution~(\ref{eq:soleta0}-\ref{eq:solw0}).
At ${\cal O}(\epsilon \delta)$ we obtain:
 \begin{align}
\partial_t \eta_1 - w_{s1} & = -  \partial_T \eta_0 +R_5^{(1)}    \, , \label{eq:eta1reduced} \\
 \partial_t w_{s1} + D\{ \eta_1 \} & = -  \partial_T w_{s0} + R_6^{(1)}    \, , \label{eq:w1reduced}
 \end{align}
 where $R_5^{(1)}$ and $R_6^{(1)}$ denote the leading, ${\cal O}(\epsilon \delta)$ contributions from $R_5$ and $R_6$. The solvability condition is obtained by forming the combination $\la e^{-i({\bf k}\cdot {\bf x}-\omega_k t)} [ -i \omega_k \text{(\ref{eq:eta1reduced})}+\text{(\ref{eq:w1reduced})} ] \ra$, which after various integrations by parts leads to:
\begin{align}
\la e^{-i({\bf k}\cdot {\bf x}-\omega_k t)} \left( - \partial_T w_{s0} +i \omega_k \partial_T \eta_0  + R_6^{(1)} -i \omega_k R_5^{(1)} \right) \ra & = 0 \, . \label{eq:reducedSC}
\end{align}
We now choose $R_5$ and $R_6$ such that this solvability condition is the same as the solvability condition~(\ref{eq:cleanerSC}) of the full 3D system. This ensures that the coefficients $\{A_{\bf k}(T)\}_{\bf k}$ entering the lowest-order solution do indeed follow the same slow dynamics as in the full system. A possible choice is $R_5=-{\bf U} \cdot \bnabla \eta$ and $R_6=-{\bf U} \cdot \bnabla w_s -\frac{1}{2} D^{-2} \{ J(\Delta \Psi, w_s)\} $, which leads to the following reduced system of 2D equations:
 \begin{align}
\partial_t \eta + {\bf U} \cdot \bnabla \eta - w_s & = 0 \, , \label{eq:reducedetaeq}\\
 \partial_t w_s + {\bf U} \cdot \bnabla w_s + D\{ \eta \} & =  \frac{1}{2} \Delta^{-1} \{ J(\Delta \Psi, w_s)\} \, . \label{eq:reducedweq}
 \end{align}
where we have used $D^{-2}=-\Delta^{-1}$.

As a side note, we remark that there are infinitely many other acceptable choices for $R_5$ and $R_6$, as the terms entering the solvability condition can be accounted for by $R_5$ or $R_6$ equivalently, and one can split these terms between $R_5$ and $R_6$ in an arbitrary fashion. The complex formulation of the next section contributes to lifting this degeneracy.

\section{Complex formulation\label{sec:remodeling}}
We expect the reduced equations to share the invariants of the original full system. Because energy is conserved in the initial system, we expect the lowest-order energy to be an adiabatic invariant of the reduced system if the background flow is time-independent. Additionally, wave action should arise as an adiabatic invariant resulting from the separation of timescales. In the spirit of the YBJ model, we now derive equivalent reduced equations that enjoy exact action and energy invariants. A priori, such equations are in no way \textit{more valid} than the initial reduced equations, but they are arguably more elegant, simpler to work with, and more amenable to numerical validation.

The first step is to introduce the complex variable associated with action conservation in the absence of background flow, $\chi=(D^{-\frac{1}{4}} \eta + i D^{-\frac{3}{4}} w_s)/\sqrt{2}$, or equivalently:
 \begin{align}
\eta & =\frac{1}{\sqrt{2}} (D^{{\frac{1}{4}}} \chi + D^{{\frac{1}{4}}} \chi^*) \, , \label{eq:etavschi}\\
w_s & =-\frac{i}{\sqrt{2}} (D^{{\frac{3}{4}}} \chi - D^{{\frac{3}{4}}} \chi^*) \, . \label{eq:wvschi}
 \end{align}
The combination $(D^{-{\frac{1}{4}}}\{$(\ref{eq:reducedetaeq})$\} + i D^{-{\frac{3}{4}}}\{$(\ref{eq:reducedweq})$\})/\sqrt{2}$ yields:
 \begin{align}
\nonumber \partial_t \chi + i D^{{\frac{1}{2}}} \chi & =  -\frac{1}{2} D^{-{\frac{1}{4}}} \{  {\bf U} \cdot \bnabla (D^{{\frac{1}{4}}} \chi) \}  -\frac{1}{2} D^{-{\frac{3}{4}}} \{  {\bf U} \cdot \bnabla (D^{{\frac{3}{4}}} \chi) \}   -\frac{1}{4} D^{-{\frac{11}{4}}} \{ J(\Delta \Psi, D^{{\frac{3}{4}}} \chi)   \} \\
&  -\frac{1}{2} D^{-{\frac{1}{4}}} \{  {\bf U} \cdot \bnabla (D^{{\frac{1}{4}}} \chi^*) \}  + \frac{1}{2} D^{-{\frac{3}{4}}} \{  {\bf U} \cdot \bnabla (D^{{\frac{3}{4}}} \chi^*) \} +\frac{1}{4} D^{-{\frac{11}{4}}} \{ J(\Delta \Psi, D^{{\frac{3}{4}}} \chi^* )\} \label{eq:step1complex}
 \end{align}
This reduced equation is designed to correctly account for the lowest-order solution and the solvability condition of the full system. Introducing an expansion of $\chi$ as $\chi=\delta [\chi_0(x,y,t,T) + \epsilon \chi_1(x,y,t,T)+\dots]$ together with the scalings~(\ref{eq:scalingBGflow}) for the background flow, the lowest-oder field $\chi_0$ satisfies:
 \begin{align}
\partial_t \chi_0 + i D^{{\frac{1}{2}}} \chi_0 & =  0 \, ,
 \end{align}
with solution:
 \begin{align}
 \chi_0 & =   \Sigma_{\bf k \in {\cal D}} B_{\bf k}(T) e^{i ({\bf k}\cdot {\bf x}-\omega_{\bf k} t)} \, . \label{eq:solchi0}
 \end{align}
We stress the fact that the complex field $\chi_0$ conveniently encodes the entire wave field in terms of complex exponentials $e^{-i \omega_{\bf k} t}$ with `positive' frequencies only, and no `negative-frequency' terms of the form $e^{+i \omega_{\bf k} t}$.
To order ${\cal O}(\epsilon \delta)$, equation~(\ref{eq:step1complex}) yields:
 \begin{align}
 \nonumber \partial_t \chi_1 + i D^{{\frac{1}{2}}} \chi_1 & =   -\frac{1}{2} D^{-{\frac{1}{4}}} \{  {\bf U}_1 \cdot \bnabla (D^{{\frac{1}{4}}} \chi_0) \}  -\frac{1}{2} D^{-{\frac{3}{4}}} \{  {\bf U}_1 \cdot \bnabla (D^{{\frac{3}{4}}} \chi_0) \}   -\frac{1}{4} D^{-{\frac{11}{4}}} \{ J(\Delta \Psi_1, D^{{\frac{3}{4}}} \chi_0)\}  \\
&  -\frac{1}{2} D^{-{\frac{1}{4}}} \{  {\bf U}_1 \cdot \bnabla (D^{{\frac{1}{4}}} \chi^*_0) \}  + \frac{1}{2} D^{-{\frac{3}{4}}} \{  {\bf U}_1 \cdot \bnabla (D^{{\frac{3}{4}}} \chi^*_0) \} +\frac{1}{4} D^{-{\frac{11}{4}}} \{ J(\Delta \Psi_1, D^{{\frac{3}{4}}} \chi_0^*) \} \, .
 \end{align}
 The solvability condition is obtained by multiplying this equation by $e^{-i({\bf k}\cdot {\bf x}-\omega_{\bf k} t)}$, before averaging over $x$, $y$ and fast time $t$. 
 Because $\chi_0^*$ involves only terms evolving as $e^{+i \omega_{\bf k} t}$ (that is, `negative frequency' oscillations), the terms involving $\chi_0^*$ do not contribute to the solvability condition. We conclude that the correct lowest-order solution and solvability condition are obtained even if one discards the $\chi^*$ terms in equation~(\ref{eq:step1complex}). We thus discard these terms to obtain a simpler reduced equation, without loss of (formal) accuracy. The new equation reads:
 \begin{align}
 \partial_t \chi + i D^{{\frac{1}{2}}} \chi & =  -\frac{1}{2} D^{-{\frac{1}{4}}} \{  {\bf U} \cdot \bnabla (D^{{\frac{1}{4}}} \chi) \}  -\frac{1}{2} D^{-{\frac{3}{4}}} \{  {\bf U} \cdot \bnabla (D^{{\frac{3}{4}}} \chi) \}   -\frac{1}{4} D^{-{\frac{11}{4}}} \{ J(\Delta \Psi, D^{{\frac{3}{4}}} \chi) \} \, .
 \end{align}
Finally, in the solvability condition we can separate the fast-time average and the space average to pack all the $D$ operators in front of each term without loss of formal accuracy. This approach is illustrated around equation~(\ref{eq:sepavg}) in appendix~\ref{app:SCzinv}. The result of this maneuver is that the contribution to the solvability condition from a term of the form $-\frac{1}{2} D^{-{\frac{1}{4}}} \{  {\bf U} \cdot \bnabla (D^{{\frac{1}{4}}} \chi) \}$ is the same as the contribution from the simpler term $-\frac{1}{2} D^{-{\frac{1}{4}}} D^{{\frac{1}{4}}} \{  {\bf U} \cdot \bnabla  \chi \} = -\frac{1}{2}  {\bf U} \cdot \bnabla  \chi $. Similarly, the contribution from the term $-\frac{1}{2} D^{-{\frac{3}{4}}} \{  {\bf U} \cdot \bnabla (D^{{\frac{3}{4}}} \chi) \}$ is the same as the contribution from the simpler term $-\frac{1}{2}   {\bf U} \cdot \bnabla \chi$. Finally, the contribution from the term $ -\frac{1}{4} D^{-{\frac{11}{4}}} \{ J(\Delta \Psi, D^{{\frac{3}{4}}} \chi)   \} $ is the same as the contribution from the simpler term $- \frac{1}{4}  D^{-1} \{ J(\Delta \Psi,  D^{-1} \chi)   \}$, where we have included part of the $D$ operators inside the Jacobian and part of them outside of the Jacobian (this splitting ensures exact action conservation in section \ref{sec:actioncons}). Making these substitutions leads to the simpler reduced evolution equation~(\ref{eq:sec2reducedBB}) for the complex variable~$\chi$, which once again shares the same leading-order solution and first solvability condition as the original full system.

\section{Narrow-band waves over a depth-dependent, 3D background flow\label{sec:NB3Dderivation}}

Having illustrated the method for the idealized situation of a depth-invariant flow, we now turn to the general case of a depth-dependent background flow. We seek a solution to the full linearized equations (\ref{eq:adimscaled13D}-\ref{eq:adimscaled43D}) using the same multiple-timescale expansion. At order $\delta$, the lowest-order solution~(\ref{eq:soleta0}-\ref{eq:solphi0}) is unchanged. However, to make progress we further assume that the waves are centered around a dimensional angular frequency $\omega_0$ with a narrow, ${\cal O}(\epsilon)$ bandwidth. To alleviate the algebra, we set the length scale $L$ used for non-dimensionalization to $L=g/\omega_0^2$. The waves are then centered around a unit dimensionless angular frequency, corresponding to a unit dimensionless wavenumber. More precisely, we assume that:
 \begin{align}
\textit{narrow-band assumption: } & A_{\bf k} = {\cal O}(1) \text{ only for } {\bf k}\in {\cal D} \text{ such that } |\omega_k- 1| = {\cal O}(\epsilon)   \label{nbassump} \\
\nonumber & A_{\bf k} \simeq 0 \text{ otherwise,}
 \end{align}
 which also corresponds to wave signal in modes with waverumber $k=1+{\cal O}(\epsilon)$ only.
At order $\epsilon \delta$, the same solvability condition~(\ref{eq:rawSC}) is obtained, with the following alternate expressions for $R_1$ through $R_4$:
 \begin{align}
R_1 & = - \partial_T \eta_0 - {\bf U}_1 \cdot \bnabla \eta_0 - (\bnabla_\perp \cdot {\bf U}_{1\perp}|_0) \eta_0 \, , \\
R_2 & = 0 \, , \\
R_3 & = - \bnabla \cdot [({{\bf u}_0} \cdot \bnabla) {\bf U}_1 + ({\bf U}_1 \cdot \bnabla) {{\bf u}_0}] \, , \label{eq:R3zdep}\\
R_4 & =  - \partial_T w_{s0} - {\bf U}_{1\perp} |_0 \cdot \bnabla_\perp w_{s0} + (\bnabla_\perp \cdot {\bf U}_{1\perp}|_0) w_{s0}  \, .
 \end{align}
 Additionally, because of the narrow-band assumption~(\ref{nbassump}) we can make the substitutions $\omega_k \rightarrow 1$ and $k \rightarrow 1$ in the expression~(\ref{eq:rawSC}) of the solvability condition without loss of formal accuracy. In appendix~\ref{app:SCzdep} we express the contribution from $R_3$ in terms of the wavy vertical velocity evaluated at the surface, the end result being:
   \begin{align}
\nonumber  \la e^{-i({\bf k}\cdot {\bf x}- t)}  \overline{e^{ {z}} (-R_3)}  \ra  & =     \la e^{-i({\bf k}\cdot {\bf x}- t)}  \left( -  \frac{1}{2 }  \Delta_\perp (\hat{\bf U}_{1\perp}) + 2  {\bf U}_{1\perp}|_0 - 2  \hat{\bf U}_{1\perp} \right) \cdot  \bnabla_\perp w_{s0} \right.  \\
  & \left. +  \left[  -\frac{1}{4} \Delta_\perp (\bnabla_\perp \cdot \hat{\bf U}_{1\perp}) - \bnabla_\perp \cdot \hat{\bf U}_{1\perp} \right] w_{s0} \ra  \, . \label{eq:R3contribution}
 \end{align}
This expression involves a weighted vertical average $ \hat{\bf U}$ of the background flow, scaled as $\hat{\bf U}=\epsilon \hat{\bf U}_1(x,y,T)$, where the $\hat{\cdot}$ operator is defined as:
\begin{align}
\hat{\cdot}=2  \, \overline{e^{2  z} (\cdot)} \, \label{eq:defUhat} \, .
\end{align}

Collecting the contributions from $R_1$, $R_2$, $R_4$ and the contribution~(\ref{eq:R3contribution}) from $R_3$ yields, after some algebra:
  \begin{align}
\nonumber 0 =  &  \la e^{-i({\bf k}\cdot {\bf x}- t)}  \left[ - \partial_T w_{s0} + i  \left[\partial_T \eta_0+ \bnabla_\perp \cdot (\eta_0 {\bf U}_{1\perp} |_0) \right] \right. \right.\\
& \left. \left. +   \bnabla_\perp \cdot (w_{s0} {\bf U}_{1\perp} |_0) -  (\bnabla_\perp \cdot \bU_1 )w_{s0} - 2  \bU_1 \cdot \bnabla_\perp w_{s0} \right]  \ra \, , \label{eq:SC3D}
 \end{align}
 up to negligible ${\cal O}(\epsilon)$ corrections. Equation~(\ref{eq:SC3D}) involves the effective horizontal flow $\bU(x,y,T)$ defined by:
\begin{align}
\bU(x,y,T) & = \hat{\bf U}_\perp + \frac{1}{4} \Delta_\perp  \hat{\bf U}_\perp \, ,\label{eq:defbU}
\end{align}
and scaled as $\bU= \epsilon \bU_1(x,y,T)$.

\subsection{Reduced 2D system}

Once again, we seek a 2D set of equations that leads to the same leading-order solution and solvability condition as the original full system under the form:
 \begin{align}
\partial_t \eta  - w_s & = R_7 \, , \\
 \partial_t w_s + D\{\eta \} & = R_8 \, ,
 \end{align}
 where the leading-order contribution from $R_7$ and $R_8$ arises at ${\cal O}(\epsilon \delta)$. Performing the same expansion as before leads to the leading-order solution (\ref{eq:soleta0}-\ref{eq:solw0}) at ${\cal O}(\delta)$,  on which we impose the narrow-band assumption~(\ref{nbassump}).
 At order $\epsilon \delta$, we obtain:
  \begin{align}
\partial_t \eta_1  - w_{s1} & = - \partial_T \eta_0+ R_7^{(1)} \, , \label{eq:eta1Oed}\\
 \partial_t w_{s1}  + D\{\eta_1 \} & = - \partial_T w_{s0} + R_8^{(1)} \, , \label{eq:phi1Oed}
 \end{align}
 where $R_7^{(1)}$ and $R_8^{(1)}$ denote the leading, ${\cal O}(\epsilon \delta)$ contributions from $R_7$ and $R_8$.
The solvability condition is obtained by forming the equation $\la e^{-i({\bf k}\cdot {\bf x}-\omega_k t)}\right.$[$-i\omega_k$(\ref{eq:eta1Oed}) $+  $(\ref{eq:phi1Oed})]$\left.   \ra $. The lhs vanishes and so must the rhs.  Without loss of formal accuracy, we can make the substitutions $\omega_k \rightarrow 1$ and $k \rightarrow 1$ in the latter based on the narrow-band assumption. This leads to:
  \begin{align}
0 = \la e^{-i({\bf k}\cdot {\bf x}- t)} (  -\partial_T w_{s0}  + i  \partial_T \eta_0 +R_8^{(1)} - i  R_7^{(1)} )  \ra \, .
 \end{align}
This solvability condition is the same as~(\ref{eq:SC3D}) if we make the following choices for $R_7$ and $R_8$:
  \begin{align}
R_7 & =  - \bnabla_\perp \cdot ( \eta {\bf U}_\perp |_0  ) \, , \\
R_8 & =  \bnabla_\perp \cdot ( w_s {\bf U}_\perp |_0  ) - w_s (\bnabla_\perp \cdot \bU ) - 2 \bU \cdot \bnabla_\perp w_s  \, ,
 \end{align}
which leads to the following set of reduced 2D equations:
 \begin{align}
 \partial_t \eta  - w_s & =  - \bnabla_\perp \cdot ( \eta {\bf U}_\perp |_0  ) \, ,  \label{eq:etazdep}\\
 \partial_t w_s + D\{\eta \} & =  \bnabla_\perp \cdot ( w_s {\bf U}_\perp |_0  ) - w_s (\bnabla_\perp \cdot \bU ) - 2 \bU \cdot \bnabla_\perp w_s \, ,  \label{eq:wzdep}
 \end{align}

\subsection{Complex variable\label{sec:complex}}

As in section~\ref{sec:remodeling}, we form the linear combination of equations $(D^{-\frac{1}{4}}\{$(\ref{eq:etazdep})$\}+i D^{-\frac{3}{4}}\{$(\ref{eq:wzdep})$\})/\sqrt{2}$ to obtain an evolution equation for the complex variable $\chi$:
 \begin{align}
\nonumber \partial_t \chi + i D^{\frac{1}{2}} \chi  = & -\frac{1}{2}D^{-\frac{1}{4}} \{ \bnabla_\perp \cdot [{\bf U}_\perp |_0 (D^{{\frac{1}{4}}} \chi +D^{{\frac{1}{4}}} \chi^*) ] \} +\frac{1}{2}D^{-\frac{3}{4}} \{ \bnabla_\perp \cdot [{\bf U}_\perp |_0 (D^{{\frac{3}{4}}} \chi -D^{{\frac{3}{4}}} \chi^*) ] \}  \\
 & - \frac{1}{2}D^{-\frac{3}{4}} \{ (\bnabla_\perp \cdot \bU) (D^{{\frac{3}{4}}} \chi -D^{{\frac{3}{4}}} \chi^*) \} - D^{-\frac{3}{4}} \{  \bU \cdot \bnabla_\perp (D^{-{\frac{3}{4}}} \chi -D^{-{\frac{3}{4}}} \chi^*)    \}  \, .
 \end{align}
 The same leading-order solution and solvability condition are obtained if we discard the $\chi^*$ terms on the rhs, which leads to the simpler equation:
 \begin{align}
\nonumber \partial_t \chi + i D^{\frac{1}{2}} \chi  = & -\frac{1}{2}D^{-\frac{1}{4}} \{ \bnabla_\perp \cdot ({\bf U}_\perp |_0 D^{{\frac{1}{4}}} \chi  ) \} +\frac{1}{2}D^{-\frac{3}{4}} \{ \bnabla_\perp \cdot ({\bf U}_\perp |_0 D^{{\frac{3}{4}}} \chi  ) \}  \\
 & - \frac{1}{2}D^{-\frac{3}{4}} \{ (\bnabla_\perp \cdot \bU) D^{{\frac{3}{4}}} \chi  \} - D^{-\frac{3}{4}} \{  \bU \cdot \bnabla_\perp (D^{{\frac{3}{4}}} \chi )    \}  \, .
 \end{align}
 Finally, we can pack all the $D$ operators in front of each term without loss of formal accuracy, as discussed at the end of section~\ref{sec:remodeling}.
This leads to the simpler reduced equation:
\begin{align}
\partial_t \chi + i D^{\frac{1}{2}} \chi   +  & \bU \cdot \bnabla_\perp \chi + \frac{1}{2} (\bnabla_\perp \cdot \bU) \chi = 0 \, . \label{eq:chi3D}
\end{align}


\subsection{Invariants}

One can easily check that equation~(\ref{eq:chi3D}) conserves the wave action ${\cal A}$. If the background flow is steady, the equation also conserves an energy invariant, identified by multiplying equation~(\ref{eq:chi3D}) by $i$, recognizing a Schr\"odinger equation, and computing $E=\la \chi^* {\cal H}\{ \chi \} \ra$ for the corresponding Hamiltonian. This leads to:
\begin{align}
\nonumber E & = \la |D^{\frac{1}{4}} \chi|^2 - i \chi^* \bU \cdot \bnabla_\perp \chi - \frac{i}{2}  (\bnabla_\perp \cdot \bU)  |\chi|^2  \ra_{\bf x} \\
& = \la |D^{\frac{1}{4}} \chi|^2 +\frac{i}{2} \left( \chi \bU \cdot \bnabla_\perp \chi^*  - \chi^* \bU \cdot \bnabla_\perp \chi \right) \ra_{\bf x}  \, .
\end{align}




\subsection{Leveraging the time-scale separation}

Equation~(\ref{eq:chi3D}) greatly reduces the computational burden of simulating surface gravity waves above by a weak background flow, because it turns an initially 3D problem into a 2D one. Additionally, as mentioned around equation~(\ref{eq:sec2reducedBB}), equation~(\ref{eq:chi3D}) can be numerically integrated using large time steps provided the fast dispersive term $i D^{{\frac{1}{2}}} \chi$ is first integrated exactly~\citep{lawson1967generalized}. We now leverage the narrow-band approximation to further highlight this point, and we derive an equation describing the evolution of the wave field over the slow advective timescale of the background flow only.
Assumption~(\ref{nbassump}) implies that, when applied to $\chi$, the operator $D^{\frac{1}{2}}$ reduces to $D^{\frac{1}{2}} = 1 + {\cal O}(\epsilon)$. After taking the fourth power we obtain $-\Delta = D^2 = 1 + {\cal O}(\epsilon)$, that is $-\Delta -1 = {\cal O}(\epsilon)$. We use this relation to expand the operator $D^{\frac{1}{2}}$ as
\begin{align}
D^{\frac{1}{2}} & =(-\Delta)^{\frac{1}{4}}=[1-\underbrace{(\Delta+1)}_{{\cal O}(\epsilon)}]^{\frac{1}{4}}= 1-\frac{1}{4} \left(1+\Delta \right) + {\cal O}(\epsilon^2)  = \frac{3}{4} - \frac{\Delta}{4} + {\cal O}(\epsilon^2)  \, ,
\end{align}
and therefore:
\begin{align}
D^{\frac{1}{2}} \chi & = \frac{3}{4} \chi - \frac{1}{4} \Delta \chi + {\cal O}(\epsilon^2 \delta) \, .
\end{align}
Substituting into the evolution equation~(\ref{eq:chi3D}) yields:
\begin{align}
\partial_t \chi + \frac{i}{4} \left( 3\chi -  \Delta \chi \right)   +  & \bU \cdot \bnabla_\perp \chi + \frac{1}{2} (\bnabla_\perp \cdot \bU) \chi = 0 \, .  \label{eq:tempchislowtime}
\end{align}
In a somewhat similar fashion to the YBJ approach, we finally introduce the demodulated complex wave amplitude ${\cal M}(x,y,t)$, defined by:
\begin{align}
{\cal M}(x,y,t)=\chi e^{i  t} \, .
\end{align}
Substituting into equation~(\ref{eq:tempchislowtime}) yields the evolution equation for ${\cal M}$:
\begin{align}
\partial_t {\cal M} - \frac{i }{4}\left( {\cal M} + \Delta {\cal M} \right)  +  & \bU \cdot \bnabla_\perp {\cal M} + \frac{1}{2} (\bnabla_\perp \cdot \bU) {\cal M} = 0 \, . \label{eq:M}
\end{align}
The second term of this equation is the dispersion term: it accounts for slight departures of the wave frequency from the unit central frequency, depending on the wavenumber. This term is ${\cal O}(\epsilon \delta)$ for narrow-band waves satisfying~(\ref{nbassump}). We conclude that ${\cal M}$ only evolves as a result of ${\cal O}(\epsilon \delta)$ terms, and equation~(\ref{eq:M}) can be integrated using large, ${\cal O}(1/\epsilon)$ timesteps.

\subsection{Eliminating the divergent part of the effective background flow\label{sec:eliminating}}

The motivation for deriving equation~(\ref{eq:M}) is not only a computational one. As shown in appendix~\ref{app:quantum}, equation~(\ref{eq:M}) also corresponds to the Schr\"odinger equation governing the motion of a charged particle in an external magnetic field (up to negligible terms). In this analogy the vector potential is proportional to $\bU$,  which can have non-zero divergence. Interestingly, quantum mechanics also indicates that the dynamics can be recast in Coulomb's gauge, for which the divergence of the vector potential vanishes. This suggests that the divergence of the background flow has a negligible effect on the wave field, and that the reduced equation can be recast in terms of a divergence-free, 2D effective background flow. We establish this result in this subsection, without referring to the quantum problem.

Consider the following Helmholtz decomposition of the background flow, $\bU= \bnabla_\perp S -\bnabla_\perp \times (\psi {\bf e}_z)$, where $S(x,y,t)={\cal O}(\epsilon)$ and $\psi(x,y,t)={\cal O}(\epsilon)$, and introduce the field:
\begin{align}
{M}(x,y,t)={\cal M}(x,y,t)e^{2iS(x,y,t)} \, . \label{eq:defcalM}
\end{align}
To highlight the scalings of the various fields and their slow time dependence, we temporarily revert to scaled notations:
\begin{align}
\psi & =\epsilon \psi_1(x,y,T) \, , \label{eq:psiscaled}\\
S & =\epsilon S_1(x,y,T) \, , \\
\bU & =\epsilon \bU_1(x,y,T) = \epsilon [\bnabla S_1(x,y,T) - \bnabla \times [\psi_1(x,y,T) {\bf e}_z] ] \, , \\
{\cal M} & =\delta {M}_0(x,y,T) e^{-2i \epsilon S_1(x,y,T)} \, .\label{eq:calMscaled}
\end{align}
where $\psi_1$, $S_1$, $\bU_1$, ${M}_0$ and their derivatives with respect to $x$, $y$ and $T$ are ${\cal O}(1)$.

The various terms in equation~(\ref{eq:M}) read:
\begin{align}
\partial_t {\cal M} & = \left[ \epsilon \delta \partial_T ({M}_0) + {\cal O}(\epsilon^2 \delta) \right] e^{-2i \epsilon S_1} \, , \\
-\frac{i}{4}({\cal M}+\Delta {\cal M}) & = -\frac{i}{4} \left[\delta ({M}_0 +  \Delta {M}_0) - \epsilon \delta [4i (\bnabla S_1)\cdot (\bnabla {M}_0) + 2i {M}_0 \Delta S_1] +{\cal O}(\epsilon^2 \delta)  \right] e^{-2i \epsilon S_1} \, ,  \\
\bU \cdot \bnabla_\perp {\cal M} & = \left[ \epsilon \delta \left[ \bnabla S_1 \cdot \bnabla {M}_0 + J(\psi_1,{M}_0) \right] + {\cal O}(\epsilon^2 \delta) \right]e^{-2i \epsilon S_1} \, , \\
\frac{1}{2} (\bnabla_\perp \cdot \bU) {\cal M} & = \epsilon \delta \frac{1}{2} (\Delta S_1) {M}_0 e^{-2i \epsilon S_1} \, ,
\end{align}
Neglecting ${\cal O}(\epsilon^2 \delta)$ terms for consistency with the previous expansion, multiplying by $e^{2i \epsilon S_1}$ and recasting the resulting equation in terms of the non-expanded variables leads to the evolution equation~(\ref{eq:sec2calMstandardadim}) for ${M}$. From the solution ${M}$ to this equation, the wave field is deduced from ${\cal M}={M} e^{-2iS}$, followed by  $\chi={\cal M}e^{-it}={M} e^{-2iS-it}$. 
Using the scaled variables~(\ref{eq:psiscaled}-\ref{eq:calMscaled}), the surface elevation is given by:
\begin{align}
\eta & =\frac{1}{\sqrt{2}} D^{\frac{1}{4}} \chi + \text{c.c.} = \frac{\delta}{\sqrt{2}} D^{\frac{1}{4}} \left\{ {M}_0(x,y,T) e^{-2i\epsilon S_1(x,y,T) -it} \right\} + \text{c.c.} \\
& = \frac{\delta}{\sqrt{2}}e^{-it} D^{\frac{1}{4}} \left\{ {M}_0(x,y,T) {e^{-2i\epsilon S_1(x,y,T)}} \right\} + \text{c.c.} \\
& = \frac{\delta}{\sqrt{2}}e^{-it} {D^{\frac{1}{4}}} \left\{ {M}_0(x,y,T)  \right\} + \text{c.c.} +{\cal O}(\epsilon \delta) \, .
\end{align}
The narrow-band assumption gives ${D^{\frac{1}{4}}} \left\{ {M}_0(x,y,T)  \right\} =  {M}_0(x,y,T) + {\cal O}(\epsilon \delta)$, so that, reverting to the unscaled variables, the leading-order surface elevation simply reads:
\begin{align}
\eta(x,y,t) & = \frac{1}{\sqrt{2}} {M}(x,y,t) e^{-it} + \text{c.c.} \, .
\end{align}
The action density is simply $|{M}|^2$ and, to leading order, the local surface-elevation variance and interfacial-slope variance averaged over one wave period are simply:
\begin{align}
\frac{1}{2 \pi} \int_t^{t+2\pi} \eta^2 \, \mathrm{d}\tilde{t} & = |{M}|^2 \, , \label{eq:dispvariance}\\
\frac{1}{2 \pi} \int_t^{t+2\pi} (\bnabla \eta)^2 \, \mathrm{d}\tilde{t} & = |\bnabla {M}|^2 \, .\label{eq:slopevariance}
\end{align}

\section{Scattering of a wave packet by a cellular flow\label{sec:scatteringanalytical}}

\subsection{Wave packet impinging on a patch of Taylor-Green flow}

\begin{figure}
    \centerline{\includegraphics[width=12 cm]{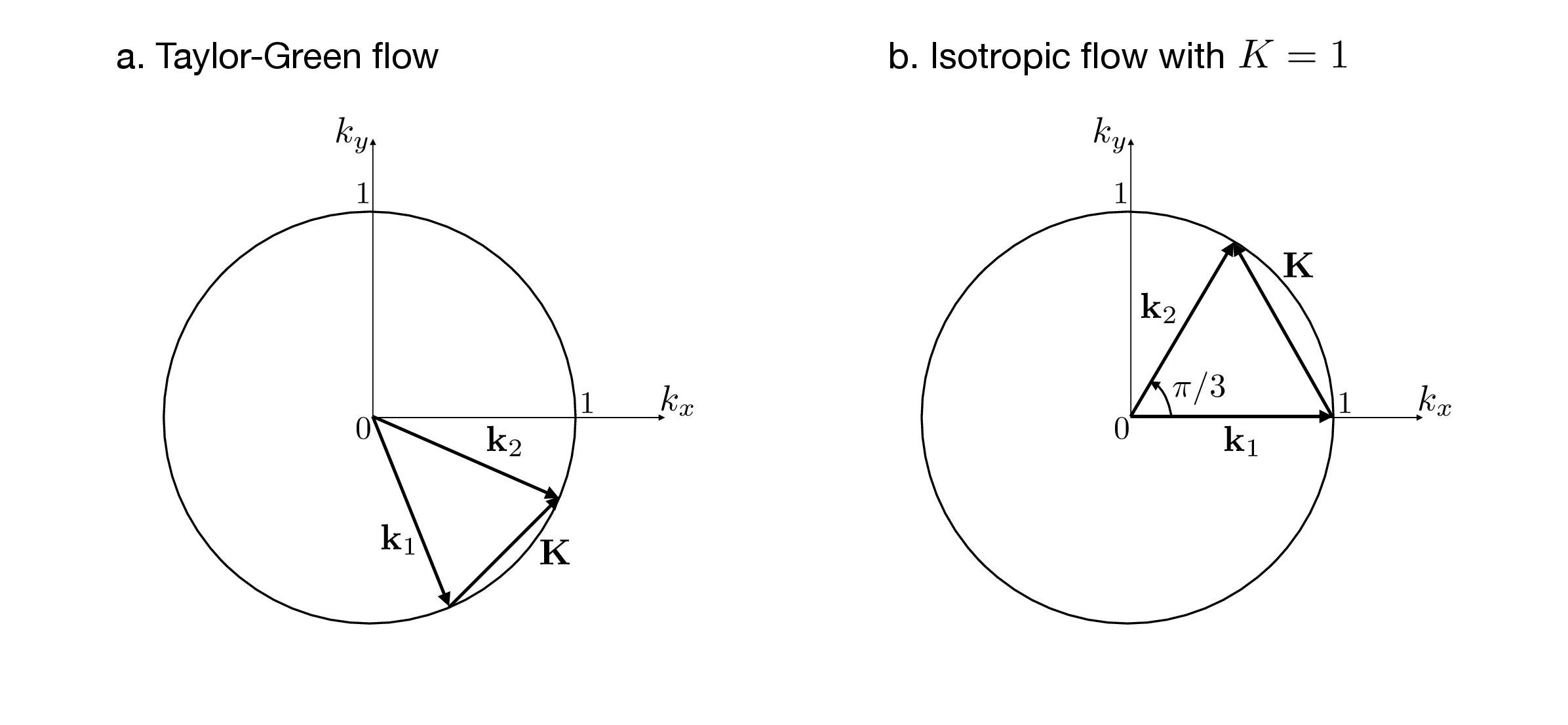} }
   \caption{\label{fig:resonantscatt} \textbf{Resonant-scattering condition.} Waves incoming with a wave vector ${\bf k}_1$ of unit norm are scattered by background flow structures with wavevector ${\bf K}$, inducing outgoing scattered waves with wave vector ${\bf k}_2={\bf k}_1+{\bf K}$. The scattering is resonant if ${\bf k}_2$ also falls on the unit circle, as illustrated  in both panels. {\bf a.} For the Taylor-Green flow in section~\ref{sec:scatteringanalytical}, the scattering wave vector ${\bf K}$ is directed along the first diagonal.  {\bf b.} For the isotropic flow in section~\ref{sec:scatteringnumerical}, incoming waves have ${\bf k}_1={\bf e}_x$. The scattering wavevector is of unit norm, $K=1$, and it can have any direction. Waves are then scattered at angles $\pm \theta_0=\pm \pi/3$ in the weak scattering regime. Only the situation corresponding to scattering at angle $+\pi/3$ is represented here.}
\end{figure}

\begin{figure}
\centering{\includegraphics[width=11 cm]{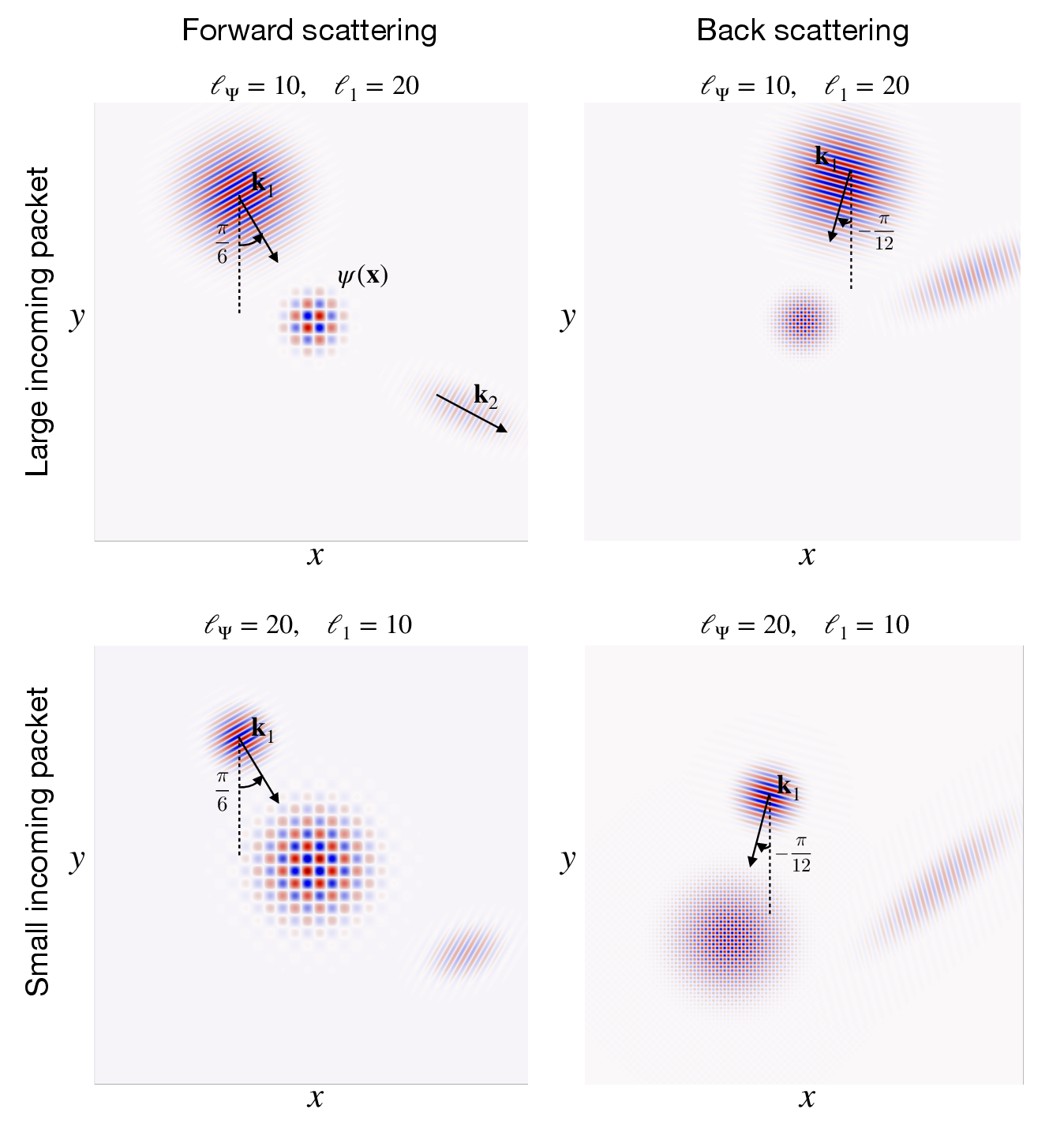} }
   \caption{Spatial structures of the incident and scattered wave packets, together with the streamfunction of the background flow. We vary the extension of the incident wave packet, the extension of the background flow, and the incidence angle. Not shown is the fact the incident wave packet keeps propagating almost undisturbed along ${\bf k}_1$ after the scattering event. \label{fig:scattering}}
\end{figure}

As an illustrative solution to the reduced equation we consider the scattering of a packet of surface waves by a localized patch of depth-invariant cellular flow, see figure~\ref{fig:scattering}. In line with equation~(\ref{eq:sec2calMstandardadim}),
time is non-dimensionalized using the frequency of the incoming waves and space is non-dimensionalized using the wavenumber of the incoming waves. The patch of steady background flow consists of 2D Taylor-Green vortices with an envelope that varies slowly in space, with the scale separation between the Taylor-Green wavelength and the size of the patch scaling as $\epsilon$. The corresponding streamfunction reads
\begin{align}
\Psi(x,y)=\epsilon \, \Xi({\bf X}) \sin(Kx/\sqrt{2}) \sin(Ky/\sqrt{2}) \, , \label{eq:TGflow}
\end{align}
where $K={\cal O}(1)$ and we have introduced the slow variable ${\bf X}=\epsilon {\bf x}$. For such a depth-invariant flow, the reduced streamfunction~$\psi$ entering equation~(\ref{eq:sec2calMstandardadim}) reads, to lowest order in $\epsilon$:
\begin{align}
\psi(x,y)=\epsilon {\psi}_1(x,y)=\epsilon \xi({\bf X}) \sin(Kx/\sqrt{2}) \sin(Ky/\sqrt{2}) \, , \qquad \text{ where }  \xi({\bf X})=\left(1-\frac{K^2}{4} \right) \Xi({\bf X}) \, . \label{eq:reducedTGflow}
\end{align}
The background flow consists of the wave vectors $(\pm K /\sqrt{2}, \pm K /\sqrt{2})$. For brevity, we consider scattering geometries involving the wavevector ${\bf K}=( K/\sqrt{2},  K/\sqrt{2})$ of the background flow only (scattering by the other three wavevectors is computed using a similar procedure). Incoming waves have wavenumber ${\bf k}_1$, with $|{\bf k}_1|=1$.
The scattering event eventually leads to free waves propagating away from the Taylor-Green patch, along the direction ${\bf k}_1+{\bf K}$. Because the background flow is steady the outgoing scattered waves have the same (unit) frequency as the incoming waves. The outgoing waves thus have unit angular frequency and wavenumber, and they are associated with the wavevector:
\begin{align}
{\bf k}_2 & =\frac{{\bf k}_1+{\bf K}}{\left|{\bf k}_1+{\bf K} \right|} \, .
\end{align}
The scattering process is efficient only close to spatial resonance, that is, when ${\bf k}_1+{\bf K}$ is almost of unit norm, such that the scattered wave satisfies the intrinsic dispersion relation. The exactly resonant situation is sketched in figure~\ref{fig:resonantscatt}a.
Allowing for a small spatial detuning, we write:
\begin{align}
{\bf k}_1+{\bf K} & =(1+\epsilon \kappa_2) {\bf k}_2 \, ,
\end{align}
where $\kappa_2={\cal O}(1)$.
We describe the scattering process using a multiple-scale expansion to the solution to equation~(\ref{eq:sec2calMstandardadim}):
\begin{align}
{M} & ={M}_0 ({\bf x},{\bf X},T) + \epsilon {M}_1 ({\bf x},{\bf X},T) + \dots \, ,
\end{align}
To ${\cal O}(1)$, equation~(\ref{eq:sec2calMstandardadim}) yields:
\begin{align}
\partial_{xx} {M}_0 + \partial_{yy} {M}_0 + {M}_0 & = 0 \, .
\end{align}
Including both the incoming and the scattered waves, the solution is:
\begin{align}
{M}_0 & = A_1({\bf X},T)e^{i {\bf k}_1 \cdot {\bf x}} + A_2({\bf X},T)e^{i {\bf k}_2 \cdot {\bf x}} \, ,
 \end{align}
 where $A_1({\bf X},T)$ and $A_2({\bf X},T)$ denote the envelopes of the incoming and scattered wave packets, respectively. To ${\cal O}(\epsilon)$, equation~(\ref{eq:sec2calMstandardadim}) yields:
\begin{align}
\frac{i}{4} \left(\partial_{xx} {M}_1 + \partial_{yy} {M}_1 + {M}_1 \right) & = \partial_T {M}_0 + J(\psi_1, {M}_0) -\frac{i}{2} \left( \partial_{Xx} {M}_0 + \partial_{Yy} {M}_0 \right) \, .
\end{align}
Two solvability conditions are obtained by demanding that the rhs contain no terms proportional to $e^{i {\bf k}_1 \cdot {\bf x}}$ or $e^{i {\bf k}_2 \cdot {\bf x}}$. When collecting such resonant terms, the advective contribution is decomposed into:
\begin{align}
J({\psi}_1, {M}_0) & = -\frac{\xi}{4\sqrt{2}} A_1 K (k_{1x}-k_{1y}) \underbrace{e^{({\bf k}_1+{\bf K}) \cdot {\bf x}}}_{e^{i {\bf k}_2 \cdot {\bf x}} e^{i \kappa_2 {\bf k}_2 \cdot {\bf X}} } -\frac{\xi}{4\sqrt{2}} A_2 K (k_{2y}-k_{2x}) \underbrace{e^{({\bf k}_2-{\bf K}) \cdot {\bf x}}}_{e^{i {\bf k}_1 \cdot {\bf x}} e^{-i \kappa_2 {\bf k}_2 \cdot {\bf X}} } + \text{N.R.} \, ,
\end{align}
where $\text{N.R.}$ stands for `non-resonant terms'. The solvability conditions finally read:
\begin{align}
\partial_T A_1 + \frac{{\bf k}_1}{2} \cdot \bnabla_{\bf X} A_1 & = - a({\bf X}) A_2 \, , \label{eq:A1}\\
\partial_T A_2 + \frac{{\bf k}_2}{2} \cdot \bnabla_{\bf X} A_2 & = a^*({\bf X}) A_1 \, ,  \label{eq:A2}
\end{align}
where $\bnabla_{\bf X}=(\partial_X,\partial_Y)$ and the function $a({\bf X})$ is defined by $a({\bf X})  = \xi({\bf X}) \frac{K}{4\sqrt{2}} (k_{1x}-k_{1y}) e^{-i \kappa_2 {\bf k}_2 \cdot {\bf X}}$, where we have used the relation $(k_{2x}, k_{2y}) = -(k_{1y}, k_{1x}) + {\cal O}(\epsilon)$ (see figure~\ref{fig:resonantscatt}).
The lhs of equations~(\ref{eq:A1}-\ref{eq:A2}) describe transport of the wave envelopes at the group velocity, while the rhs of these equations describe conversion of incoming waves into scattered waves and vice-versa.

\subsection{Weak-scattering limit}

We now solve equations~(\ref{eq:A1}-\ref{eq:A2}) in the limit of a weak scattering process, for which $A_2 \ll A_1$. Accordingly, we neglect the rhs of equation~(\ref{eq:A1}), which reduces to $\partial_T A_1 + \frac{{\bf k}_1}{2} \cdot \bnabla_{\bf X} A_1=0$. We solve this equation by changing variables from $({\bf X},T)$ to $({\bf X}'={\bf X}-\frac{{\bf k}_1}{2} T,\tau=T)$, which simply leads to:
\begin{align}
A_1({\bf X},T)=f \left({\bf X}-\frac{{\bf k}_1 T }{2} \right) \, ,
\end{align}
where the function $f$ denotes the (prescribed) shape of the incoming wave packet. This shape propagates at the group velocity without any deformation (dispersion only arises at next order in scale separation). To compute the scattered wave packet long after the scattering event, we insert this expression in the rhs of~(\ref{eq:A2}), before changing variables from $({\bf X},T)$ to $({\bf X}''={\bf X}-\frac{{\bf k}_2}{2} T,\tau=T)$ and integrating from $\tau=-\infty$ to $\tau=+\infty$. This leads to:
\begin{align}
A_2({\bf X}'')=\int_{-\infty}^{+\infty} a^* \left( {\bf X}'' + \frac{{\bf k}_2}{2} \tau \right) f\left(  {\bf X}'' + \frac{{\bf k}_2-{\bf k}_1}{2} \tau \right)  \mathrm{d}\tau \, .  \label{eq:A2structure}
\end{align}
This expression provides the envelope of the scattered wave packet, in the frame propagating with the group velocity ${\bf k}_2/2$ of the scattered waves. The integral on the rhs is easily evaluated when the envelopes of both the incoming wave packet and the patch of cellular flow are Gaussian. We consider this situation in the following, denoting as $\ell_1$ the spatial extent of the incoming Gaussian wave packet and as $\ell_\Psi$ that of the patch of background flow. We focus on the regime where these dimensionless scales are large, scaling as $1/\epsilon$, and introduce $L_1=\epsilon \ell_1 = {\cal O}(1)$ and $L_\Psi=\epsilon \ell_\Psi = {\cal O}(1)$. We thus write the structures of the incoming wave packet and patch of background flow as:
\begin{align}
f({\bf X}) & =f_{\text{max}} \exp \left( -\frac{{\bf X}^2}{2 L_1^2}\right) \, , \\
\Xi({\bf X}) & =\Xi_{\text{max}}  \exp \left( -\frac{{\bf X}^2}{2 L_\Psi^2}\right) \, .
\end{align}
Evaluating the integral on the rhs of~(\ref{eq:A2structure}) leads to the following structure for the modulus of the scattered wave packet:
\begin{align}
|A_2({\bf X}'')| =|f_{\text{max}}| \Xi_{\text{max}} K  \left(1-\frac{K^2}{4} \right) \left| k_{1x}-k_{1y} \right| \frac{\sqrt{\pi}}{2\sqrt{\frac{1}{L_\Psi^2}+\frac{K^2}{L_1^2}}} \exp \left( -\frac{\kappa_2^2}{\frac{2}{L_\Psi^2}+\frac{2K^2}{L_1^2}} \right) G({\bf X}'') \, , \label{eq:modA2}
\end{align}
where the Gaussian function $G({\bf X}'')$ has maximum value one and is given by:
\begin{align}
G({\bf X}'') & = \exp \left\{ - \frac{{\bf X}''^2}{2} \left(\frac{1}{L_\Psi^2}+\frac{1}{L_1^2} \right) +\frac{\left[{\bf X}'' \cdot \left[ \frac{{\bf k}_1}{ L_\Psi^2}  +   \left( \frac{1}{ L_\Psi^2}+\frac{1}{ L_1^2}  \right) {\bf K}    \right]   \right]^2}{\frac{2}{L_\Psi^2}+\frac{2 K^2}{ L_1^2}}\right\} \, .
\end{align}
As expected, expression~(\ref{eq:modA2}) indicates that the scattered wave amplitude is proportional to both the amplitude of the incoming wave and the typical velocity $K \Xi_{\text{max}}$ of the background flow. The factor $\left(1-\frac{K^2}{4} \right)$ arises from the difference between the true background flow $\Psi$ and the reduced background flow $\psi$ entering equation~(\ref{eq:sec2calMstandardadim}). This factor smoothens the transition to zero as $K$ reaches $2$, which corresponds to the maximum value of $K$ for which resonance can occur. For $K>2$, $|{\bf k}_1 + {\bf K}|$ differs from one, $\kappa_2$ becomes significant and the exponential factor $\exp \left[ -{\kappa_2^2}/\left({\frac{2}{L_\Psi^2}+\frac{2 K^2}{L_1^2}}\right) \right]$
makes the scattered amplitude tiny (the flow is then transparent to the incoming waves). More generally, this exponential factor characterizes the width of the resonance, showing that scattering is significantly suppressed as soon as 
$||{\bf k}_1 + {\bf K}|-1| \, \min\{ \ell_\Psi,\ell_1 \} \gtrsim 1$. The spatial resonance condition is thus more peaked when a long wave packet gets scattered by an extended patch of cellular flow.

In figure~\ref{fig:scattering} we represent the incoming and scattered wave packets, together with the streamfunction of the scattering background flow (using arbitrary amplitudes). We consider various angles of incidence, including both forward-scattering and back-scattering geometries. In each case the characteristic wavenumber $K$ of the Taylor-Green array of vortices has been chosen to achieve exact spatial resonance, with $|{\bf k}_1 + {\bf K}|=1$  (see figure~\ref{fig:resonantscatt}a).

\section{Scattering of a wave packet by a disorganized flow\label{sec:scatteringnumerical}}

\begin{figure}
    \centerline{\includegraphics[width=15 cm]{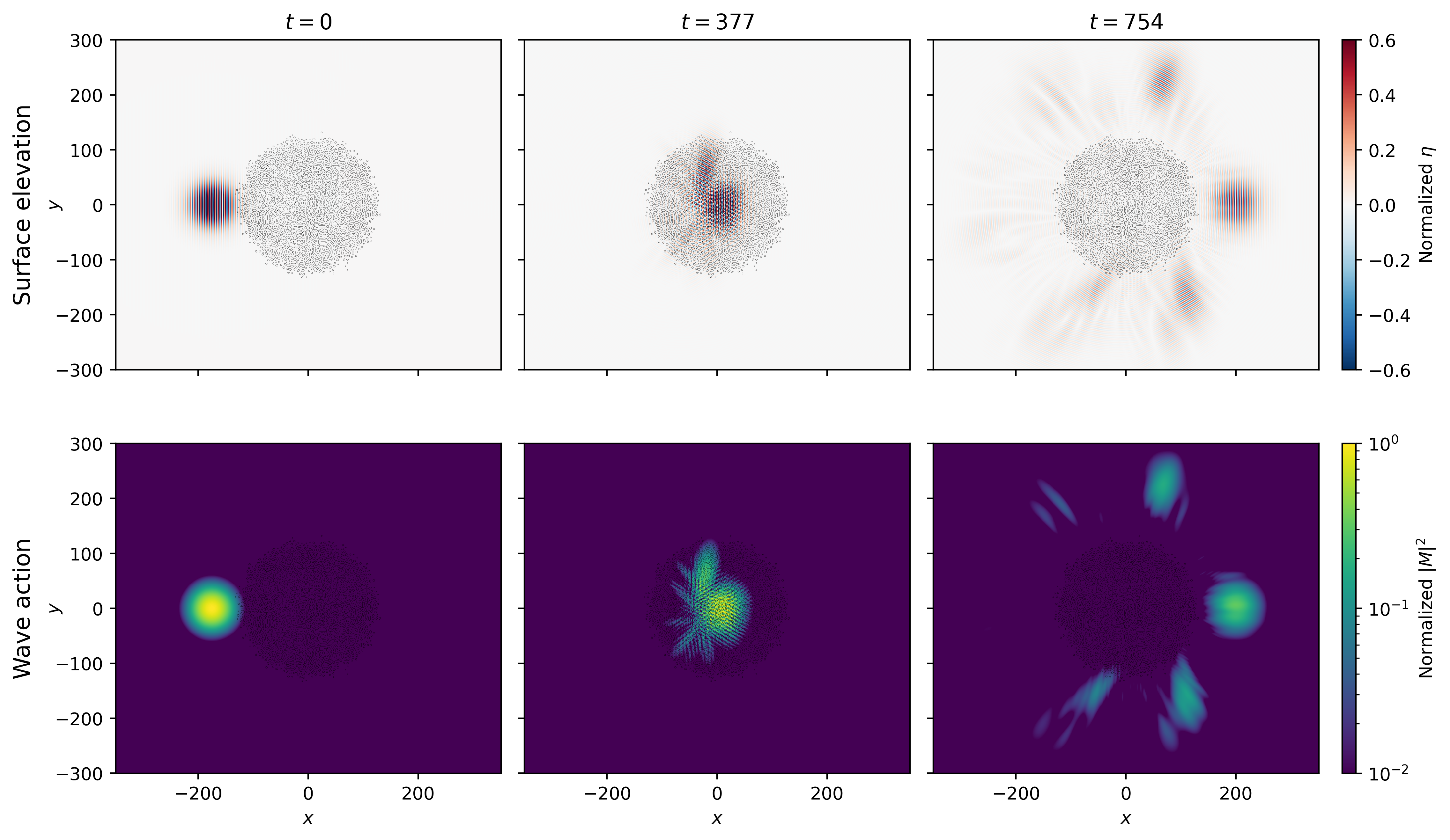} }
   \caption{\label{fig:snapshotsnum} Scattering of a Gaussian wave packet by a patch of disorganized flow. The patch of flow has a Gaussian envelope and corresponds to random complex Fourier amplitudes over a ring of wave vectors of norm $K=1$ (see text). The  Froude number based on the rms velocity of the flow is $0.18$ in the central region. 
   Incoming waves (left) interact with the patch of background flow (center), which induces scattered waves (right). The first scattering angles $\pm \theta_0 = \pm \pi/3$ are clearly visible in the bottom-right panel, together with the secondary scattering angles $\pm 2 \theta_0 = \pm 2 \pi/3$. Once it leaves the scattering region, the initial wave packet has lost a significant fraction of its initial action. 
   The surface elevation $\eta$ (top), and the wave action $|M|^2$ (bottom) are normalized using their maximum values at $t=0$.}
\end{figure}

To further illustrate the capabilities of the reduced equations, we turn to stronger wave scattering by a patch of disorganized flow, which we address by numerically solving equation~(\ref{eq:sec2calMstandardadim}). The initial condition is a rightward-propagating wave packet of Gaussian envelope, see figure~\ref{fig:snapshotsnum}. 
A disorganized, steady effective streamfunction $\psi(x,y)$ is synthesized by considering nonzero Fourier complex coefficients in a narrow ring of wave vectors with $K\simeq 1$ only, with random phases. 
The Froude number based on the rms velocity of the resulting flow is set to $0.09$ (rms value of the dimensionless velocity). This streamfunction is then multiplied by a Gaussian envelope of dimensionless standard deviation 44, which leads to the local patch of disorganized flow visible in figure~\ref{fig:snapshotsnum}.

Having set the effective background flow and the wave initial condition, we solve equation~(\ref{eq:sec2calMstandardadim}) using a standard pseudo-spectral method with {an Integrating Factor RK4 time-stepper~\citep{lawson1967generalized} to exactly integrate the free-propagation operator $i(\Delta+1)/4$. In figure~\ref{fig:snapshotsnum}, we report the resulting spatial distributions of surface-wave displacement $\eta(x,y,t)$ and wave action $|{M}|^2$, normalized by their maximum value in the initial state. The incoming wave packet (left-hand panel) travels rightward and gets distorted as it overlaps with the background flow (central panel). Eventually, waves are scattered in various directions (right-hand panel). For weak scattering, the preferential scattering direction $\theta_0$ -- relative to the direction ${\bf e}_x$ of propagation of the incoming waves -- is given by the spatial resonance condition: the sum of the wave vector ${\bf k}_1={\bf e}_x$ of the incoming waves and of a scattering wave vector of norm $K=1$ from the background flow must give a wave vector ${\bf k}_2$ of unit norm for the scattered waves. As illustrated in figure~\ref{fig:resonantscatt}b, this can only happen for the scattering angles $\pm \theta_0$ with $\theta_0= 2 \arcsin \frac{K}{2}=\pi/3$. Beyond such weak scattering, multiple scattering is also visible in figure~\ref{fig:snapshotsnum}: the scattered waves in turn get scattered, leading to waves escaping the patch of background flow at angles $\pm 2 \theta_0=\pm 2 \pi /3$. Even weaker waves also escape the scattering region in other angular directions. Also visible is the fact that the initial wave packet has lost a significant fraction of its action once it leaves the scattering region.

\section{Conclusion and perspectives}

We have computed reduced, 2D equations governing the evolution of deep-water surface gravity waves above a slow background flow. We have illustrated the possibilities offered by these equations by computing the scattering of a wave packet impinging on a background flow, analytically in the weak-scattering regime and numerically for stronger scattering by a disorganized background flow. Incoming waves with wavenumber $k$ are scattered by background-flow structures of wavenumber $2k$ at most, depending on relative orientation, while smaller-scale background flows are transparent to such waves (a result consistent with previous studies focused on the Born scattering regime, see~\citet{phillips1959scattering} and~\citet{fabrikant1994influence}).

The reduced equations were derived in two limiting situations: broad-band waves above a depth-invariant background flow and narrow-band waves above a fully 3D background flow. In the latter situation, the reason for focussing on narrow-band wave fields is that the effective horizontal flow $\bU$ explicitly depends on the horizontal wavenumber -- or, equivalently, the wave frequency -- through the weighted vertical average $\int_{-\infty}^0 e^{2kz} {\bf U} \, \mathrm{d}z$ of the background flow ${\bf U}$, where $k$ denotes the horizontal wavenumber  ($k$ is absorbed into the non-dimensionalization of section~\ref{sec:NB3D} and therefore it is absent from expression~(\ref{eq:defUhat})).
A wave field with narrow-band signal around several central frequencies could easily be dealt with by separating it into its narrow-band frequency components. Each frequency component then obeys an equation~(\ref{eq:sec2calMstandardadim}), with an effective background streamfunction ${\psi}$ that explicitly depends on the value of the central frequency. After numerical or analytical integration of the equations, the full solution is obtained by superposition. Alternatively, an evolution equation for broad-band waves above a fully 3D flow could be obtained if the vertical dependence of the flow were prescribed, and simple enough that $\int_{-\infty}^0 e^{2kz} {\bf U} \, \mathrm{d}z$ could be expressed explicitly in terms of $k$. This would lead to an equation of the form~(\ref{eq:chi3D}) for broad-band waves, with an effective horizontal flow $\bU$ that explicitly involves a function of the operator $D$.

To construct the reduced equations we have introduced an equivalent solvability condition method. While the present study focuses on deep-water surface gravity waves, the approach could readily be extended to other surface-wave systems, such as finite-depth gravity-capillary waves, possibly including forcing and dissipation. The present method also suggests ways of including wave-wave nonlinearities. Indeed, the lhs of equations (\ref{eq:2Deqeta}-\ref{eq:2Deqw}) correspond to the linear part of the HOS system designed to simulate weakly nonlinear surface waves~\citep{dommermuth1987high,west1987new,onorato2002freely,dyachenko2004weak,zhang2022numerical,higgins2024numerical}. Including the standard nonlinear terms of the HOS system back into equations~(\ref{eq:reducedetaeq}-\ref{eq:reducedweq}) readily leads to  a reduced system of 2D equations governing the evolution of broad-band nonlinearly interacting waves above a depth-invariant background flow:
\begin{align}
\begin{split}\label{eq:HOS_eta}
    \partial_t \eta + {\bf U} \cdot \bnabla \eta - w_s ={}& -\bnabla \cdot \left( \eta \bnabla D^{-1} w_s \right) - D \{ \eta w_s \} + D\left\{ \eta D\{ \eta w_s\} \right\} \\
         & - \frac{1}{2}D \left\{ \eta^2 Dw_s \right\} + \frac{1}{2} \Delta(\eta^2 w_s) \, ,
\end{split}\\
\begin{split}\label{eq:HOS_w}
    \partial_t w_s + {\bf U} \cdot \bnabla w_s + D\eta ={}& \frac{1}{2} \Delta^{-1} \{ J(\Delta \Psi, w_s)\} - \frac{1}{2} D \left\{ (\bnabla D^{-1} w_s)^2 - w_s^2 \right\} \\
         & - D \left\{ w_s  D\{\eta w_s\} \right\} + D \left\{ \eta w_s  D w_s\right\} \, . 
\end{split}
\end{align}

This system provides a way to study the influence of a background flow on the nonlinear dynamics of interacting surface gravity waves, including in the weak turbulence regime~\citep{Nazarenko2011,FalconMordant2022}.

Finally, another source of nonlinearity is the feedback of the wave field onto the background flow. Including this feedback in a reduced set of equations would lead to a two-way-coupled model between the waves and the mean flow. In an oceanographic context, such a model would be key to study the equilibrated regime of Langmuir cells, where the emergent cellular flow feeds back onto the waves that induce them. In a very recent preprint, \citet{onuki2026reduced} report precisely on such a two-way coupled model, combining the Craik-Leibovich equation with a reduced equation obtained through reconstitution in the narrow-band regime. The reduced wave equations proposed in the present work offer a more compact formulation, explicitly highlighting the central role of the vertical vorticity of the background flow while demonstrating the negligible contributions from horizontal vorticity and near-surface divergence. In a companion paper~\citep{galletpart2}, we leverage this minimalistic form to derive a compact coupled model between waves and background flow, including two-way coupling and wave-wave interactions in a consistent fashion.

\medskip
\noindent  \textbf{Funding:} This research is supported by the European Research Council under grant agreement 101124590.

\medskip
\noindent \textbf{Competing interests:} The authors declare none.

\appendix

\section{Solvability condition in terms of surface quantities only\label{app:SC}}

\subsection{The case of a depth-independent background flow\label{app:SCzinv}}

Here is the procedure to express the terms involving  $R_3$ in the solvability condition~(\ref{eq:rawSC})  in terms of wavy quantities evaluated at the surface only.
Firstly, using the fact that ${\bf u}_0$ is irrotational we have:
 \begin{align}
\bnabla \cdot [({{\bf u}_0} \cdot \bnabla) {\bf U}_1 + ({\bf U}_1 \cdot \bnabla) {{\bf u}_0}] & = \Delta ({\bf U}_1\cdot {\bf u}_0)-{\bf u}_0 \cdot \Delta {\bf U}_1 \, .
 \end{align}
(Note that this equality also holds in the case where ${\bf U}$ is fully 3D, with three non-vanishing components). Secondly, only the resonant part $R_3^{\text{(res)}}$ of $R_3$ (proportional to $e^{-i \omega_k t}$) enters the solvability condition~(\ref{eq:rawSC}). Because the background flow is depth invariant, $R_3^{\text{(res)}}$ has the same vertical dependence as ${\bf u}_0$, namely it is proportional to $e^{kz}$. This argument is based on the frequency resonance condition only, which imposes a vertical dependence in $e^{k z}$ for the lowest-order solution at frequency $\omega_k$. We thus obtain:
\begin{align}
\la e^{-i({\bf k}\cdot {\bf x}-\omega_k t)} \overline{ e^{k {z}} R_3}\ra & =  \la e^{-i({\bf k}\cdot {\bf x}-\omega_k t)}  \overline{ e^{k {z}} R_3^{\text{(res)}} } \ra  \\
& = \la e^{-i({\bf k}\cdot {\bf x}-\omega_k t)}   \overline{ e^{2 k {z}} R_3^{\text{(res)}}|_0 }  \ra  \\
& =  \la e^{-i({\bf k}\cdot {\bf x}-\omega_k t)} \frac{R_3|_0}{2 k} \ra  \\
& =  \la e^{-i({\bf k}\cdot {\bf x}-\omega_k t)} \frac{1}{2 k} (-\Delta ({\bf U}_1\cdot {\bf u}_0)|_0+{\bf u}_0|_0 \cdot \Delta {\bf U}_1) \ra \\
& =  \la e^{-i({\bf k}\cdot {\bf x}-\omega_k t)} \frac{1}{2 k} {\bf u}_0|_0 \cdot \Delta {\bf U}_1 \ra \\
& =  \la e^{-i({\bf k}\cdot {\bf x}-\omega_k t)} \frac{1}{2 k} \bnabla (\phi|_0) \cdot [-\bnabla \times (\Delta \psi_1 {\bf e}_z)] \ra \\
& =  \la e^{-i({\bf k}\cdot {\bf x}-\omega_k t)} \frac{1}{2} D^{-1} \{ J(\Delta \psi_1, D^{-1}{w_{s0}})\}  \ra \, ,
\end{align}
where we have used the relation $\phi|_0=D^{-1}{w_{s0}}$ to obtain the last equality. 

Further simplification is obtained by separating the fast-time average $\la \cdot \ra_t$ and the space average $\la \cdot \ra_{\bf x}$, using the fact that the background flow is independent of the fast time $t$:
\begin{align}
\la e^{-i({\bf k}\cdot {\bf x}-\omega_k t)}   \overline{ e^{k {z}} R_3 }) \ra & =   \la e^{-i {\bf k}\cdot {\bf x}} \frac{1}{2} D^{-1} \{ J(\Delta \psi_1, D^{-1}\{  \la e^{i \omega_k t} w_{s0} \ra_{t} \})\}  \ra_{\bf x} \, . \label{eq:sepavg}
\end{align}
From~(\ref{eq:solw0}) we know that $\la e^{i \omega_k t} w_{s0} \ra_{t}$ only involves wavenumber $k$, so that $D^{-1}\{  \la e^{i \omega_k t} w_{s0} \ra_{t} \} =  \la e^{i \omega_k t} w_{s0} \ra_{t} /k$. Then:
\begin{align}
\la e^{-i({\bf k}\cdot {\bf x}-\omega_k t)} \overline{ e^{k {z}} R_3 } \ra & =   \frac{1}{2k} \la e^{-i {\bf k}\cdot {\bf x}}  D^{-1} \{ J(\Delta \psi_1,   \la e^{i \omega_k t} w_{s0} \ra_{t} )\}  \ra_{\bf x} \, .
\end{align}
Using $\la e^{-i {\bf k}\cdot {\bf x}}  D^{-1} \{ \cdot \}\ra_{\bf x}=\la e^{-i {\bf k}\cdot {\bf x}}   ( \cdot ) \ra_{\bf x}/k$ (an application of the formula $\la f^* D^{\alpha} \{g \} \ra_{\bf x} = \la D^{\alpha} \{ f^* \} g \ra_{\bf x}$, easily obtained by expanding $f$ and $g$ in Fourier series) finally yields:
\begin{align}
\la e^{-i({\bf k}\cdot {\bf x}-\omega_k t)} \overline{ e^{k {z}} R_3 }  \ra & =   \frac{1}{2k^2} \la e^{-i {\bf k}\cdot {\bf x}}   J(\Delta \psi_1,   \la e^{i \omega_k t} w_{s0} \ra_{t} )  \ra_{\bf x} \, \\
& =  \frac{1}{2k^2} \la e^{-i ({\bf k}\cdot {\bf x}-\omega_k t)}   J(\Delta \psi_1,   w_{s0}  )  \ra \, .
\end{align}

\subsection{The case of a depth-dependent background flow\label{app:SCzdep}}
 
Here we evaluate the contribution to the solvability condition from the $z$-integral of $R_3$, when the background flow is fully 3D. We first notice that, because ${\bf u}_0$ is irrotational, 
 \begin{align}
- R_3 & = \Delta ({\bf U}_1\cdot {\bf u}_0) - {\bf u}_0 \cdot \Delta {\bf U}_1 \\
& = \partial_{zz}  ({\bf U}_1\cdot {\bf u}_0) + \Delta_\perp ({\bf U}_1\cdot {\bf u}_0) -   {\bf u}_0 \cdot \Delta_\perp {\bf U}_1 - {\bf u}_0 \cdot \partial_{zz} {\bf U}_1  \, .
 \end{align}
The sought integral is:
 \begin{align}
 \la e^{-i({\bf k}\cdot {\bf x}- t)}  \overline{e^{ {z}} (-R_3)}  \ra  = &   \la e^{-i({\bf k}\cdot {\bf x}- t)}  \left[ \overline{e^{ {z}} \partial_{zz}  ({\bf U}_1\cdot {\bf u}_0)} - k^2  \overline{e^{ {z}}  {\bf U}_1\cdot {\bf u}_0} \right. \right. \label{eq:tempR3} \\
\nonumber & \left. \left.  - \overline{e^{ {z}}   (\Delta_\perp {\bf U}_1) \cdot {\bf u}_0} - \overline{e^{ {z}} \partial_{zz}  ({\bf U}_1) \cdot {\bf u}_0} \right]   \ra  \, . 
 \end{align}
 Introducing the shorthand notation $\hat{\bf U}=2  \overline{e^{2  z} {\bf U}}$ and leveraging the $e^{ z}$ vertical dependence of the resonant part of ${\bf u}_0$, after several integrations by parts in $z$ the terms inside the square bracket can be replaced by (equality for the resonant part only):
\begin{align}
  \overline{e^{ {z}} \partial_{zz}  ({\bf U}_1\cdot {\bf u}_0)}  & \rightarrow \left(\partial_z ({\bf U}_1)|_0 +\frac{1}{2} \hat{\bf U}_1 \right) \cdot {\bf u}_0 |_0 \, , \\
  - k^2  \overline{e^{ {z}}  {\bf U}_1\cdot {\bf u}_0}  & \rightarrow - \frac{1}{2} \hat{\bf U}_1 \cdot {\bf u}_0 |_0 \, , \\
  - \overline{e^{ {z}}   (\Delta_\perp {\bf U}_1) \cdot {\bf u}_0} & \rightarrow -\frac{1}{2 }  \Delta_\perp (\hat{\bf U}_1) \cdot {\bf u}_0 |_0 \, , \\
  - \overline{e^{ {z}} \partial_{zz}  ({\bf U}_1) \cdot {\bf u}_0} & \rightarrow \left( -\partial_z ({\bf U}_1)|_0 + 2  {\bf U}_1|_0 - 2  \hat{\bf U}_1 \right)\cdot {\bf u}_0 |_0  \, .
\end{align}
Substituting into~(\ref{eq:tempR3}) leads to:
  \begin{align}
 \la e^{-i({\bf k}\cdot {\bf x}- t)}  \overline{e^{ {z}} (-R_3)}  \ra  = &   \la e^{-i({\bf k}\cdot {\bf x}- t)}  \left(-  \frac{1}{2 }  \Delta_\perp (\hat{\bf U}_1) + 2  {\bf U}_1|_0 - 2 \hat{\bf U}_1 \right) \cdot {\bf u}_0 |_0  \ra  \, . \label{eq:tempR3bis}
 \end{align}
One can check that this contribution reduces to the expression in the previous section when the background flow is independent of $z$ and divergence-free.
We further simplify the expression above using the fact that ${\bf u}_0$ is a potential flow that satisfies ${\bf u}_0|_0=\bnabla_\perp \phi|_0 +  \phi|_0 {\bf e}_z + {\cal O}(\epsilon\delta)$ under the narrow-band assumption. We also make use of the fact that the vertical component of $\hat{\bf U}_1$ satisfies $\hat{W}_1=(\bnabla_\perp \cdot \hat{\bf U}_{1\perp})/2 $, as easily deduced from the incompressibility constraint.
Substituting into~(\ref{eq:tempR3bis}) and using the boundary condition $W_1|_0=0$ yields:
  \begin{align}
\nonumber  \la e^{-i({\bf k}\cdot {\bf x}- t)}  \overline{e^{ {z}} (-R_3)}  \ra  & =   \la e^{-i({\bf k}\cdot {\bf x}- t)}  \left( -  \frac{1}{2 }  \Delta_\perp (\hat{\bf U}_{1\perp}) + 2  {\bf U}_{1\perp}|_0 - 2 \hat{\bf U}_{1\perp} \right) \cdot  \bnabla_\perp \phi|_0  \right.  \\
  & \left. +  \left[  -\frac{1}{4 } \Delta_\perp (\bnabla_\perp \cdot \hat{\bf U}_{1\perp}) - \bnabla_\perp \cdot \hat{\bf U}_{1\perp} \right]    \phi|_0 \ra  \, . \label{eq:tempR3ter} \\
  & =   \la e^{-i({\bf k}\cdot {\bf x}- t)}  \left( -  \frac{1}{2 }  \Delta_\perp (\hat{\bf U}_{1\perp}) + 2  {\bf U}_{1\perp}|_0 - 2  \hat{\bf U}_{1\perp} \right) \cdot  \bnabla_\perp w_{s0} \right.  \\
  & \left. +  \left[  -\frac{1}{4 } \Delta_\perp (\bnabla_\perp \cdot \hat{\bf U}_{1\perp}) - \bnabla_\perp \cdot \hat{\bf U}_{1\perp} \right] w_{s0} \ra  \, ,
 \end{align}
 where we substituted $\phi|_0 = w_{s0} + {\cal O}(\epsilon \delta)$ to obtain the last equality, which holds up to negligible, ${\cal O}(\epsilon \delta)$ corrections.



\section{Quantum analogy and the use of Coulomb's gauge\label{app:quantum}}

{Throughout this appendix only we change non-dimensionalization by introducing the variables (watch out for the factors $1/2$ and $2$):
\begin{align}
 {\bf x}=\check{\bf x} \, , \qquad \frac{t}{2} =\check{t} \, , \qquad  2 \bU=\check{\bU} \, .
\end{align}
After dropping the $\check{\cdot}$ for brevity, equation~(\ref{eq:M}) reduces to:
\begin{align}
\partial_t {\cal M} - \frac{i }{2}\left( {\cal M} + \Delta {\cal M} \right)  +  & \bU \cdot \bnabla_\perp {\cal M} + \frac{1}{2}(\bnabla_\perp \cdot \bU) {\cal M} = 0 \, . \label{eq:Mter}
\end{align}
Mutliplying by $i$ leads to the following Schr\"odinger form for the equation:
\begin{align}
i \partial_t {\cal M} & = {\cal H} \{ {\cal M} \} \, , \\
{\cal H} \{ {\cal M} \} & = \frac{1}{2} (-i \bnabla +\bU)^2 - \frac{1+\bU^2}{2} \, .
\end{align}
The term $-\bU^2/2$ from the Hamiltonian leads to a term of order $\epsilon^2 \delta$ in the evolution equation, which we neglect for consistency with the asymptotic expansion, leading to:
\begin{align}
i \partial_t {\cal M} & = {\cal H} \{ {\cal M} \} \, , \\
{\cal H} \{ {\cal M} \} & = \frac{1}{2} (-i \bnabla +\bU)^2 - \frac{1}{2} \, .
\end{align}
We recognize the Hamiltonian for a particle of unit mass and unit charge in a magnetic field, using units such that $\hbar=1$. The vector potential ${\bf A}$ is given by ${\bf A}  = - \bU$.
Below are a few remarks regarding this analogy:
\begin{itemize}
\item Of course, one can absorb the constant term $-1/2$ from the Hamiltonian into an oscillatory phase in time, but then one must be careful to restrict attention to particles with initial energy of order $-1/2+{\cal O}(\epsilon)$ to stay within the narrow-band assumption~(\ref{nbassump}) under which the equation was derived.
\item Upon expanding the term $\frac{1}{2} (-i \bnabla +\bU)^2$ a negligible contribution $\bU^2/2$ arises. Yet, we keep the full term $\frac{1}{2} (-i \bnabla +\bU)^2$ in the Hamiltonian to highlight the analogy with particles in a magnetic field. 
\item Finally, we note that the vector potential ${\bf A}$ has non-zero divergence, because $\bnabla \cdot \bU \neq 0$. The quantum analogy suggests that changing gauge to the Coulomb gauge $\bnabla \cdot {\bf A}={\bf 0}$ could provide an effective divergence-free flow advecting the waves. This is the rationale behind section~\ref{sec:eliminating} of the main text.
\end{itemize}


\section{Scale-separation limit, large-scale flows and uniform shear flows\label{app:scaleseparation}}

While the goal of the present study is to characterize the interaction of waves and mean-flow with comparable scales, we can check that we recover the expected behavior of the system in the more familiar limit of scale separation. 

\subsection{Scale-separation for broad-band waves over a depth-invariant flow}

When the waves are short compared to the scale of the background flow, we have $\Delta \psi \ll \psi$, recalling that space is non-dimensionalized with the typical wavelength over $2\pi$. The effect of the background flow then simplifies to $J(\psi, \chi) + D^{-1} J(\Delta\psi, D^{-1} \chi) \simeq J(\psi, \chi)$, and equation~\eqref{eq:sec2reducedBB} reduces to:


   \begin{align}
  \partial_t \chi + i D^{{\frac{1}{2}}} \chi +  J(\psi,\chi ) &  = 0 \, . \label{eq:sec2scaleseparationlimit}
 \end{align}
 This equation is the gravity-wave version of the equation we considered in \citet{tlili2026equilibrium} to simulate capillary waves above a background flow in the regime of scale separation.
Inserting a plane-wave ansatz $\chi \propto e^{-i \Omega t + i {\bf k}\cdot {\bf x}}$ into equation~(\ref{eq:sec2scaleseparationlimit}) readily gives the standard Doppler-shifted dispersion relation $\Omega({\bf x},{\bf k})=\sqrt{k}+{\bf U}\cdot {\bf k}$. Additionally, in the scale separation limit the last term in the energy invariant~(\ref{eq:sec2energyinvariant}) is negligible as compared to the middle one, for the latter contains more derivatives of the wave field and fewer derivatives of the background flow. The energy invariant becomes:
   \begin{align}
E & \simeq \la |D^{{\frac{1}{4}}} \chi|^2  - i \psi J(\chi,\chi^*) \ra_{\bf x} =  \la |D^{{\frac{1}{4}}} \chi|^2  - i \chi^* J(\psi,\chi) \ra_{\bf x} =  \la |D^{{\frac{1}{4}}} \chi|^2  - i \chi^* {\bf U}\cdot \bnabla \chi \ra_{\bf x} \\
\nonumber & \simeq \la |\chi|^2 ( \sqrt{k}  + {\bf U}\cdot {\bf k} ) \ra_{\bf x} \, ,
 \end{align}
 where we have inserted the plane-wave ansatz to obtain the last equality.
We recover the standard result that the absolute energy density equals the action density $|\chi|^2$ multiplied by the absolute frequency $\Omega({\bf x},{\bf k})$.


\subsection{Including a horizontally uniform vertically sheared background flow\label{app:uniform}}

From equation~(\ref{eq:defbU}), the horizontally uniform part of the effective background flow $\bU$ is simply obtained as $\la \bU \ra_{\bf x}=\la \hat{{\bf U}} \ra_{\bf x}$. Such a horizontally uniform background flow is non-divergent and can easily be included in the lhs of equation~(\ref{eq:sec2calMstandardadim}) as an additional advective contribution  $\la \hat{{\bf U}} \ra_{\bf x} \cdot \bnabla {M}$.
As an example, consider the situation where the background flow is a horizontally invariant vertically sheared flow, ${\bf U}=[U(z),V(z),0]$. Equation~(\ref{eq:sec2calMstandardadim})  augmented with the term $\la \hat{{\bf U}} \ra_{\bf x} \cdot \bnabla {M}$ reduces to:
\begin{align}
\partial_t {M} + \la \hat{{\bf U}} \ra_{\bf x} \cdot \bnabla {M} - \frac{i}{4} \left( \Delta {M} + {M} \right) & = 0 \, , \label{eq:Mmeanflow}
\end{align}
and, because the effective flow $ \la \hat{{\bf U}} \ra_{\bf x}$ is horizontally non-divergent, the definition of ${M}$ reduces to ${M}={\cal M}=\chi e^{it}$. An ansatz of the form $\chi \propto e^{i ({\bf k}\cdot {\bf x}-\Omega t)}$ corresponds to ${M} \propto e^{i ({\bf k}\cdot {\bf x}-(\Omega-1) t)}$, and after substitution in~(\ref{eq:Mmeanflow}) one obtains the following dispersion relation:
\begin{align}
\Omega & = 1+\frac{k^2-1}{4} +  \la \hat{{\bf U}} \ra_{\bf x} \cdot {\bf k} \, .
\end{align}
The first two terms correspond to the expansion of the intrinsic dispersion relation of deep-water surface gravity waves near $k^2=1$, valid in the narrow-band regime: $\sqrt{k}=(k^2)^{1/4}=[1+(k^2-1)]^{1/4}\simeq 1+ (k^2-1)/4$.
The last term corresponds to the standard correction $\la \hat{{\bf U}} \ra_{\bf x}$  to the phase velocity of the waves, as discussed e.g. by~\citet{stewart1974hf} and~\citet{kirby1989surface}.

\subsection{Scale-separation for narrow-band waves over a depth-dependent flow\label{app:scalesep3D}}

In the main text we assume that the background flow has a characteristic horizontal scale comparable to the wave length. In this appendix, we consider the scale-separation regime where the background flow has a horizontal scale much greater than the wavelength. In that case, instead of integrating $\bnabla S = {\cal O}(\epsilon)$ as $S=\epsilon S_1(x,y,T)$, we integrate it as $S=S_0(\epsilon x,\epsilon y,T)$, where $S_0$ evolves slowly in space, over a dimensionless horizontal scale $1/\epsilon$. As compared to equations (\ref{eq:vaguedefcalM}), (\ref{eq:sec2etavscalM}) and (\ref{eq:sec2wvscalM}) in the main text, which are valid in the absence of scale separation only, the variables ${M}$, $\eta$ and $w_s$ inherit slowly evolving phase factors $e^{\pm 2i S_0}$ in the scale-separated regime, with:
\begin{align}
{M} & = \chi e^{it} e^{2 i S_0}  \, , \label{eq:calMSS}\\
\eta(x,y,t) & = \frac{1}{\sqrt{2}} {M}(x,y,t)e^{-it} e^{-2 i S_0} + \text{c.c.} \, , \label{eq:etaSS} \\
w_s(x,y,t) & = -\frac{i}{\sqrt{2}} {M}(x,y,t) e^{-it} e^{-2 i S_0} + \text{c.c.}   \, . \label{eq:wSS}
\end{align}
Consider a plane-wave ansatz for $\chi$ under the form $\chi \propto e^{i({\bf k}\cdot {\bf x}-\Omega t)}$, corresponding to ${M} \propto e^{i[{\bf k}\cdot {\bf x}-(\Omega-1) t]-2 i S_0}$. Leveraging the slow evolution of $S_0$ in space, we recast ${M}$ under the local plane-wave form ${M} \propto e^{i [ ({\bf k}-2 \bnabla S_0) \cdot {\bf x}-(\Omega-1) t ]}$, where $ \bnabla S_0={\cal O}(\epsilon)$. We substitute the latter plane-wave form into the reduced equation~(\ref{eq:sec2calMstandardadim}), whose advective term is recast as $-[\bnabla \times (\psi {\bf e}_z)] \cdot \bnabla {M}$.  After neglecting terms of order $\epsilon^2$, the resulting dispersion relation reads:
\begin{align}
\Omega & = 1+\frac{k^2-1}{4} + [\bnabla S - \bnabla \times (\psi {\bf e}_z)] \cdot {\bf k} \, .
\end{align}
We recognize the expanded version of the intrinsic dispersion relation in the narrow-band regime, together with a Doppler-shift term associated with the flow $\bU$. For the large-horizontal-scale background flow considered here, the relation~(\ref{eq:defbU}) reduces to $\bU = \hat{\bf U}_\perp$, so that the dispersion relation becomes
\begin{align}
\Omega & = 1+\frac{k^2-1}{4} + \hat{\bf U}_\perp \cdot {\bf k} \, .
\end{align}
We recover again the standard form of the Doppler-shift term, as discussed by~\citet{stewart1974hf} and~\citet{kirby1989surface}.
The additional phases $e^{\pm 2i S_0}$ entering the relations (\ref{eq:calMSS}-\ref{eq:wSS}) should be included in the predictions if accurate phase information is needed. However, accurate phase information is perhaps rarely needed in the regime of scale separation. To predict the wave strength, one only needs to use equation~(\ref{eq:sec2calMstandardadim}) to predict the wave action density $|{M}|^2$, also equal to twice the variance of surface elevation averaged over one period, see equation~(\ref{eq:dispvariance}). This clearly shows that the divergence of the effective background flow does not affect these quantities at leading order, be it in the presence or in the absence of scale separation. 


\bibliographystyle{jfm}

\bibliography{SSW}

\begin{thebibliography}{52}
\expandafter\ifx\csname natexlab\endcsname\relax\def\natexlab#1{#1}\fi
\def\au#1{#1} \def\ed#1{#1} \def\yr#1{#1}\def\at#1{#1}\def\jt#1{\textit{#1}}
  \def\bt#1{#1}\def\bvol#1{\textbf{#1}} \def\vol#1{#1} \def\pg#1{#1}
  \def\publ#1{#1}\def\arxiv#1{#1}\def\org#1{#1}\def\st#1{\textit{#1}}

\bibitem[Berthet {\em et~al.\/}(2003)Berthet, Fauve \&
  Labb{\'e}]{berthet2003study}
{\sc \au{Berthet, R}, \au{Fauve, S} \& \au{Labb{\'e}, R}} \yr{2003}  \at{Study
  of the sound-vortex interaction: direct numerical simulations and
  experimental results}.  \jt{The European Physical Journal B-Condensed Matter
  and Complex Systems}  \bvol{32}~(2),  \pg{237--242}.

\bibitem[Berthet \& Lund(1995)]{berthet1995forward}
{\sc \au{Berthet, R{\'e}my} \& \au{Lund, Fernando}} \yr{1995}  \at{The forward
  scattering of sound by vorticity}.  \jt{Physics of Fluids}  \bvol{7}~(10),
  \pg{2522--2524}.

\bibitem[B{\^o}as \& Young(2020)]{boas2020directional}
{\sc \au{B{\^o}as, Ana B~Villas} \& \au{Young, William~R}} \yr{2020}
  \at{Directional diffusion of surface gravity wave action by ocean
  macroturbulence}.  \jt{Journal of Fluid Mechanics}  \bvol{890},  \pg{R3}.

\bibitem[Boury {\em et~al.\/}(2023)Boury, Bühler \& Shatah]{Boury2023}
{\sc \au{Boury, S.}, \au{Bühler, O.} \& \au{Shatah, J.}} \yr{2023}
  \at{Fast-slow wave transitions induced by a random mean flow}.  \jt{Physical
  Review E}  \bvol{108}~(5),  \pg{055101}.

\bibitem[Bühler(2014)]{buhler2014}
{\sc \au{Bühler, Oliver}} \yr{2014} {\em Waves and Mean Flows\/}, 2nd edn.
  \publ{Cambridge University Press}.

\bibitem[Cerda \& Lund(1993)]{cerda1993interaction}
{\sc \au{Cerda, Enrique} \& \au{Lund, Fernando}} \yr{1993}  \at{Interaction of
  surface waves with vorticity in shallow water}.  \jt{Physical review letters}
   \bvol{70}~(25),  \pg{3896}.

\bibitem[Chini {\em et~al.\/}(2014)Chini, Malecha \& Dreeben]{chini2014large}
{\sc \au{Chini, GP}, \au{Malecha, Z} \& \au{Dreeben, TD}} \yr{2014}
  \at{Large-amplitude acoustic streaming}.  \jt{Journal of fluid mechanics}
  \bvol{744},  \pg{329--351}.

\bibitem[Dommermuth \& Yue(1987)]{dommermuth1987high}
{\sc \au{Dommermuth, Douglas~G} \& \au{Yue, Dick~KP}} \yr{1987}  \at{A
  high-order spectral method for the study of nonlinear gravity waves}.
  \jt{Journal of Fluid Mechanics}  \bvol{184},  \pg{267--288}.

\bibitem[Dyachenko {\em et~al.\/}(2004)Dyachenko, Korotkevich \&
  Zakharov]{dyachenko2004weak}
{\sc \au{Dyachenko, Alexander~I}, \au{Korotkevich, Alexander~O} \&
  \au{Zakharov, Vladimir~E}} \yr{2004}  \at{Weak turbulent kolmogorov spectrum
  for surface gravity waves}.  \jt{Physical review letters}  \bvol{92}~(13),
  \pg{134501}.

\bibitem[Fabrikant \& Raevsky(1994)]{fabrikant1994influence}
{\sc \au{Fabrikant, AL} \& \au{Raevsky, MA}} \yr{1994}  \at{The influence of
  drift flow turbulence on surface gravity wave propagation}.  \jt{Journal of
  Fluid Mechanics}  \bvol{262},  \pg{141--156}.

\bibitem[Falcon \& Mordant(2022)]{FalconMordant2022}
{\sc \au{Falcon, Eric} \& \au{Mordant, Nicolas}} \yr{2022}  \at{Experiments in
  surface gravity–capillary wave turbulence}.  \jt{Annual Review of Fluid
  Mechanics}  \bvol{54}~(1),  \pg{1--25}.

\bibitem[Gallet(2026)]{galletpart2}
{\sc \au{Gallet, Basile}} \yr{2026}  \at{Surface gravity wave–mean flow
  interaction with comparable spatial scales. part 2: two-way coupling and
  wave-wave interactions.}  \jt{submitted to Journal of Fluid Mechanics} .

\bibitem[Gallet \& Young(2014)]{gallet2014refraction}
{\sc \au{Gallet, Basile} \& \au{Young, William~R}} \yr{2014}  \at{Refraction of
  swell by surface currents}.  \jt{Journal of Marine Research}  \bvol{72},
  \pg{105--126}.

\bibitem[Guti{\'e}rrez \& Auma{\^\i}tre(2016)]{gutierrez2016surface}
{\sc \au{Guti{\'e}rrez, Pablo} \& \au{Auma{\^\i}tre, S{\'e}bastien}} \yr{2016}
  \at{Surface waves propagating on a turbulent flow}.  \jt{Physics of Fluids}
  \bvol{28}~(2).

\bibitem[Higgins \& Gallet(2024)]{higgins2024numerical}
{\sc \au{Higgins, Christopher} \& \au{Gallet, Basile}} \yr{2024}  \at{Numerical
  validation of the inverse cascade of surface gravity wave action}.
  \jt{Physical Review Letters}  \bvol{132}~(16),  \pg{164002}.

\bibitem[Humbert {\em et~al.\/}(2017{\natexlab{{\em a\/}}})Humbert, Aumaitre \&
  Gallet]{humbert2017surface}
{\sc \au{Humbert, Thomas}, \au{Aumaitre, S} \& \au{Gallet, B}}
  \yr{2017{\natexlab{{\em a\/}}}}  \at{Surface-wave doppler velocimetry in a
  liquid metal: Inferring the bifurcations of the subsurface flow}.
  \jt{Europhysics Letters}  \bvol{119}~(2),  \pg{24001}.

\bibitem[Humbert {\em et~al.\/}(2017{\natexlab{{\em b\/}}})Humbert,
  Auma{\^\i}tre \& Gallet]{humbert2017wave}
{\sc \au{Humbert, Thomas}, \au{Auma{\^\i}tre, S{\'e}bastien} \& \au{Gallet,
  Basile}} \yr{2017{\natexlab{{\em b\/}}}}  \at{Wave-induced vortex recoil and
  nonlinear refraction}.  \jt{Physical Review Fluids}  \bvol{2}~(9),
  \pg{094701}.

\bibitem[Kirby \& Chen(1989)]{kirby1989surface}
{\sc \au{Kirby, James~T} \& \au{Chen, Tsung-Muh}} \yr{1989}  \at{Surface waves
  on vertically sheared flows: approximate dispersion relations}.  \jt{Journal
  of Geophysical Research: Oceans}  \bvol{94}~(C1),  \pg{1013--1027}.

\bibitem[Kunze(1985)]{kunze1985near}
{\sc \au{Kunze, Eric}} \yr{1985}  \at{Near-inertial wave propagation in
  geostrophic shear}.  \jt{Journal of Physical Oceanography}  \bvol{15}~(5),
  \pg{544--565}.

\bibitem[Labb{\'e} \& Pinton(1998)]{labbe1998propagation}
{\sc \au{Labb{\'e}, R} \& \au{Pinton, J-F}} \yr{1998}  \at{Propagation of sound
  through a turbulent vortex}.  \jt{Physical review letters}  \bvol{81}~(7),
  \pg{1413}.

\bibitem[Lawson(1967)]{lawson1967generalized}
{\sc \au{Lawson, J~Douglas}} \yr{1967}  \at{Generalized runge-kutta processes
  for stable systems with large lipschitz constants}.  \jt{SIAM Journal on
  Numerical Analysis}  \bvol{4}~(3),  \pg{372--380}.

\bibitem[Lighthill(1978)]{lighthill1978acoustic}
{\sc \au{Lighthill, James}} \yr{1978}  \at{Acoustic streaming}.  \jt{Journal of
  sound and vibration}  \bvol{61}~(3),  \pg{391--418}.

\bibitem[Lund \& Rojas(1989)]{lund1989ultrasound}
{\sc \au{Lund, Fernando} \& \au{Rojas, Cristian}} \yr{1989}  \at{Ultrasound as
  a probe of turbulence}.  \jt{Physica D: nonlinear phenomena}
  \bvol{37}~(1-3),  \pg{508--514}.

\bibitem[Michel \& Chini(2019)]{michel2019strong}
{\sc \au{Michel, Guillaume} \& \au{Chini, Gregory~P}} \yr{2019}  \at{Strong
  wave--mean-flow coupling in baroclinic acoustic streaming}.  \jt{Journal of
  Fluid Mechanics}  \bvol{858},  \pg{536--564}.

\bibitem[Nazarenko(2011)]{Nazarenko2011}
{\sc \au{Nazarenko, Sergey}} \yr{2011} {\em Wave Turbulence\/}.  \publ{Springer
  Berlin Heidelberg}.

\bibitem[Nyborg(1958)]{nyborg1958acoustic}
{\sc \au{Nyborg, Wesley~L}} \yr{1958}  \at{Acoustic streaming near a boundary}.
   \jt{The Journal of the Acoustical Society of America}  \bvol{30}~(4),
  \pg{329--339}.

\bibitem[Onorato {\em et~al.\/}(2002)Onorato, Osborne, Serio, Resio, Pushkarev,
  Zakharov \& Brandini]{onorato2002freely}
{\sc \au{Onorato, Miguel}, \au{Osborne, Alfred~Richard}, \au{Serio, Marina},
  \au{Resio, D}, \au{Pushkarev, A}, \au{Zakharov, Vladimir~E} \& \au{Brandini,
  C}} \yr{2002}  \at{Freely decaying weak turbulence for sea surface gravity
  waves}.  \jt{Physical review letters}  \bvol{89}~(14),  \pg{144501}.

\bibitem[Onuki \& Fujiwara(2026)]{onuki2026reduced}
{\sc \au{Onuki, Yohei} \& \au{Fujiwara, Yasushi}} \yr{2026}  \at{A reduced
  model for surface wave-current interactions without spatial scale
  separation}.  \jt{arXiv preprint arXiv:2606.03231} .

\bibitem[Phillips(1959)]{phillips1959scattering}
{\sc \au{Phillips, OM}} \yr{1959}  \at{The scattering of gravity waves by
  turbulence}.  \jt{Journal of Fluid Mechanics}  \bvol{5}~(2),  \pg{177--192}.

\bibitem[Prabhudesai {\em et~al.\/}(2022)Prabhudesai, Perrard, Petrelis \&
  Fauve]{prabhudesai2022statistics}
{\sc \au{Prabhudesai, Gaurav}, \au{Perrard, St{\'e}phane}, \au{Petrelis,
  Francois} \& \au{Fauve, Stephan}} \yr{2022}  \at{Statistics of phase
  fluctuations of an acoustic wave propagating through a turbulent flow}.
  \jt{Europhysics Letters}  \bvol{140}~(4),  \pg{43001}.

\bibitem[Rayleigh(1896)]{rayleigh1896theory}
{\sc \au{Rayleigh, John William Strutt~Baron}} \yr{1896} {\em The theory of
  sound\/}, ,  \vol{vol.~2}.  \publ{Macmillan}.

\bibitem[Roberts(1985)]{roberts1985introduction}
{\sc \au{Roberts, AJ}} \yr{1985}  \at{An introduction to the technique of
  reconstitution}.  \jt{SIAM journal on mathematical analysis}  \bvol{16}~(6),
  \pg{1243--1257}.

\bibitem[Stewart \& Joy(1974)]{stewart1974hf}
{\sc \au{Stewart, Robert~H} \& \au{Joy, Joseph~W}} \yr{1974} Hf radio
  measurements of surface currents.  \bt{In {\em Deep sea research and
  oceanographic abstracts\/}}, ,  \vol{vol.~21},  \pg{pp. 1039--1049}.
  Elsevier.

\bibitem[Teixeira \& Belcher(2002)]{teixeira2002distortion}
{\sc \au{Teixeira, MAC} \& \au{Belcher, SE}} \yr{2002}  \at{On the distortion
  of turbulence by a progressive surface wave}.  \jt{Journal of Fluid
  Mechanics}  \bvol{458},  \pg{229--267}.

\bibitem[Thomas(2017)]{thomas2017new}
{\sc \au{Thomas, Jim}} \yr{2017}  \at{New model for acoustic waves propagating
  through a vortical flow}.  \jt{Journal of Fluid Mechanics}  \bvol{823},
  \pg{658--674}.

\bibitem[Thomas \& Yamada(2018)]{thomas2018amplitude}
{\sc \au{Thomas, Jim} \& \au{Yamada, Ray}} \yr{2018}  \at{An amplitude equation
  for surface gravity wave-topography interactions}.  \jt{Physical Review
  Fluids}  \bvol{3}~(12),  \pg{124802}.

\bibitem[Tlili \& Gallet(2025)]{Tlili2025}
{\sc \au{Tlili, Alexandre} \& \au{Gallet, Basile}} \yr{2025}  \at{Statistics of
  near-inertial waves over a background flow via quantum and statistical
  mechanics}.  \jt{submitted to Journal of Fluid Mechanics} .

\bibitem[Tlili \& Gallet(2026)]{tlili2026equilibrium}
{\sc \au{Tlili, Alexandre} \& \au{Gallet, Basile}} \yr{2026}  \at{Equilibrium
  statistical mechanics of waves in inhomogeneous moving media}.  \jt{Physical
  Review Letters}  \bvol{136}~(24),  \pg{244001}.

\bibitem[Vanneste \& Young(2026)]{vanneste2026consistent}
{\sc \au{Vanneste, Jacques} \& \au{Young, William~R}} \yr{2026}  \at{A
  consistent phase-averaged model of the interactions between surface gravity
  waves and currents}.  \jt{arXiv preprint arXiv:2602.21976} .

\bibitem[Vincent {\em et~al.\/}(2025)Vincent, Henry, Kumar, Botton,
  Poth{\'e}rat \& Miralles]{vincent2025phenomenology}
{\sc \au{Vincent, Bjarne}, \au{Henry, Daniel}, \au{Kumar, Abhishek},
  \au{Botton, Val{\'e}ry}, \au{Poth{\'e}rat, Alban} \& \au{Miralles, Sophie}}
  \yr{2025}  \at{Phenomenology of laminar acoustic streaming jets}.
  \jt{Physical Review Fluids}  \bvol{10}~(11),  \pg{114103}.

\bibitem[Vincent {\em et~al.\/}(2024)Vincent, Miralles, Henry, Botton \&
  Poth{\'e}rat]{vincent2024experimental}
{\sc \au{Vincent, Bjarne}, \au{Miralles, Sophie}, \au{Henry, Daniel},
  \au{Botton, Val{\'e}ry} \& \au{Poth{\'e}rat, Alban}} \yr{2024}
  \at{Experimental study of a helical acoustic streaming flow}.  \jt{Physical
  Review Fluids}  \bvol{9}~(2),  \pg{024101}.

\bibitem[Vivanco \& Melo(2000)]{vivanco2000surface}
{\sc \au{Vivanco, Francisco} \& \au{Melo, Francisco}} \yr{2000}  \at{Surface
  spiral waves in a filamentary vortex}.  \jt{Physical Review Letters}
  \bvol{85}~(10),  \pg{2116}.

\bibitem[Vivanco \& Melo(2004)]{vivanco2004experimental}
{\sc \au{Vivanco, Francisco} \& \au{Melo, Francisco}} \yr{2004}
  \at{Experimental study of surface waves scattering by a single vortex and a
  vortex dipole}.  \jt{Physical Review E}  \bvol{69}~(2),  \pg{026307}.

\bibitem[Wagner {\em et~al.\/}(2017)Wagner, Ferrando \&
  Young]{wagner2017asymptotic}
{\sc \au{Wagner, Gregory~L}, \au{Ferrando, Gwenael} \& \au{Young, William~R}}
  \yr{2017}  \at{An asymptotic model for the propagation of oceanic internal
  tides through quasi-geostrophic flow}.  \jt{Journal of Fluid Mechanics}
  \bvol{828},  \pg{779--811}.

\bibitem[Wang {\em et~al.\/}(2023)Wang, B{\^o}as, Young \&
  Vanneste]{wang2023scattering}
{\sc \au{Wang, Han}, \au{B{\^o}as, Ana B~Villas}, \au{Young, William~R} \&
  \au{Vanneste, Jacques}} \yr{2023}  \at{Scattering of swell by currents}.
  \jt{Journal of Fluid Mechanics}  \bvol{975},  \pg{A1}.

\bibitem[Wang {\em et~al.\/}(2025)Wang, Villas~Bôas, Vanneste \&
  Young]{Wang2025}
{\sc \au{Wang, Han}, \au{Villas~Bôas, Ana~B.}, \au{Vanneste, Jacques} \&
  \au{Young, William~R.}} \yr{2025}  \at{Scattering of surface waves by ocean
  currents: the u2h map}.  \jt{Journal of Fluid Mechanics}  \bvol{1005},
  \pg{A12}.

\bibitem[West {\em et~al.\/}(1987)West, Brueckner, Janda, Milder \&
  Milton]{west1987new}
{\sc \au{West, Bruce~J}, \au{Brueckner, Keith~A}, \au{Janda, Ralph~S},
  \au{Milder, D~Michael} \& \au{Milton, Robert~L}} \yr{1987}  \at{A new
  numerical method for surface hydrodynamics}.  \jt{Journal of Geophysical
  Research: Oceans}  \bvol{92}~(C11),  \pg{11803--11824}.

\bibitem[Young \& {Ben Jelloul}(1997)]{YBJ1997}
{\sc \au{Young, W.R.} \& \au{{Ben Jelloul}, Mahdi}} \yr{1997}  \at{Propagation
  of near-inertial oscillations through a geostrophic flow}.  \jt{Journal of
  Marine Research}  \bvol{55}~(4).

\bibitem[Zakharov {\em et~al.\/}(1992)Zakharov, L'vov \& Falkovich]{zlf_book}
{\sc \au{Zakharov, V.E.}, \au{L'vov, V.S.} \& \au{Falkovich, G.}} \yr{1992}
  {\em Kolmogorov Spectra of Turbulence I\/}.  \publ{Springer Berlin,
  Heidelberg}.

\bibitem[Zakharov \& Zaslavskii(1982)]{zakharov_1982}
{\sc \au{Zakharov, V.E.} \& \au{Zaslavskii, M.M.}} \yr{1982}  \at{The kinematic
  equation and kolmogorov spectra in a weak turbulence theory of wind waves}.
  \jt{Akademiia Nauk SSSR, Izvestiia, Fizika Atmosfery i Okeana}  \bvol{18},
  \pg{970--979}.

\bibitem[Zhang \& Pan(2022)]{zhang2022numerical}
{\sc \au{Zhang, Zhou} \& \au{Pan, Yulin}} \yr{2022}  \at{Numerical
  investigation of turbulence of surface gravity waves}.  \jt{Journal of Fluid
  Mechanics}  \bvol{933},  \pg{A58}.

\bibitem[Zuccoli {\em et~al.\/}(2025)Zuccoli, Brambley \&
  Barkley]{zuccoli2025deep}
{\sc \au{Zuccoli, Emanuele}, \au{Brambley, Edward~J} \& \au{Barkley, Dwight}}
  \yr{2025}  \at{Deep-water closure model for surface waves on axisymmetric
  swirling flows}.  \jt{Physical Review Fluids}  \bvol{10}~(2),  \pg{024801}.

\end{thebibliography}

\end{document}